\documentclass[9.5pt,journal,compsoc]{IEEEtran}
\IEEEoverridecommandlockouts
\usepackage{cite}
\usepackage[colorlinks=true,linkcolor=cyan,citecolor=blue]{hyperref}
\usepackage{amsmath,amssymb,amsfonts}
\usepackage{amsthm}
\usepackage{algorithmic}
\usepackage{graphicx}
\usepackage{textcomp}
\usepackage{dsfont}
\usepackage{xcolor}
\usepackage{float}
\newtheorem{theorem}{Theorem}
\newtheorem{rem}{Remark}

\newtheorem{defn}{Definition}
\newtheorem*{Remark*}{Remark}
\def\BibTeX{{\rm B\kern-.05em{\sc i\kern-.025em b}\kern-.08em
    T\kern-.1667em\lower.7ex\hbox{E}\kern-.125emX}}
    
    \makeatletter
\newcommand{\linebreakand}{%
  \end{@IEEEauthorhalign}
  \hfill\mbox{}\par
  \mbox{}\hfill\begin{@IEEEauthorhalign}
}

\makeatother

 \usepackage{bbm, subfig}
\usepackage{multirow}

\newcommand{\bX}{\mathbf{X}}
\newcommand{\bY}{\textbf{Y}}
\newcommand{\bV}{\mathbf{V}}

\newcommand{\bR}{\mathbf{R}}

\newcommand{\bP}{\mathbf{P}}

\newcommand{\bA}{\mathbf{A}}
\newcommand{\bW}{\mathbf{W}}
\newcommand{\bS}{\mathbf{S}}

\newcommand{\bD}{\mathbf{D}}

\newcommand{\bZ}{\mathbf{Z}}
\newcommand{\bI}{\mathbf{I}}

\newcommand{\bU}{\mathbf{U}}
\newcommand{\RR}{\mathbb{R}}
\newcommand{\bVci}{\mathbf{V}^{(c)}_{i}}

\newcommand{\bVc}{\mathbf{V}^{(c)}}
\newcommand{\bVct}{(\mathbf{V}^{(c)})^\top}
\newcommand{\bAc}{\mathbf{A}^{(c)}}
\newcommand{\bXc}{\mathbf{X}^{(c)}}
\newcommand{\bXct}{(\mathbf{X}^{(c)})^\top}
\newcommand{\hbXc}{\widehat{\mathbf{X}}^{(c)}}
\newcommand{\hbX}{\widehat{\mathbf{X}}}
\newcommand{\tbXc}{\widetilde{\mathbf{X}}^{(c)}}
\newcommand{\bPc}{\mathbf{P}^{(c)}}
\newcommand{\bZc}{\mathbf{Z}^{(c)}}
\newcommand{\bSigma}{\boldsymbol \Sigma}
\newcommand{\whp}{\text{w.h.p.}}

\theoremstyle{definition}
\theoremstyle{definition}
\theoremstyle{definition}
\theoremstyle{definition}
\theoremstyle{definition}
\theoremstyle{definition}
\newtheorem{assumption}{Assumption}%[section]

\begin{document}
\title{Detection of Model-based Planted Pseudo-cliques in Random Dot Product Graphs by the Adjacency Spectral Embedding and the Graph Encoder Embedding}

\author{Tong Qi, Vince Lyzinski% <-

\markboth{Journal of \LaTeX\ Class Files,~Vol.~14, No.~8, August~2015}%
{Shell \MakeLowercase{\textit{et al.}}: Bare Demo of IEEEtran.cls for IEEE Journals}

\IEEEcompsocitemizethanks{
\IEEEcompsocthanksitem T. Qi is with the Department of Mathematics, University of Maryland, College Park, MD. E-mail: \href{mailto:tqi@umd.edu}{tqi@umd.edu}.

\IEEEcompsocthanksitem V. Lyzinski is with the Department of Mathematics, University of Maryland, College Park, MD. E-mail: \href{mailto:vlyzinsk@umd.edu}{vlyzinsk@umd.edu}.

\IEEEcompsocthanksitem This material is based on research sponsored by the Air Force Research
Laboratory (AFRL) and Defense Advanced Research Projects Agency (DARPA) under agreement number
FA8750-20-2-1001. 
The U.S. Government is authorized to reproduce and distribute reprints for Governmental purposes
notwithstanding any copyright notation thereon. The views and conclusions contained herein are
those of the authors and should not be interpreted as necessarily representing the official policies
or endorsements, either expressed or implied, of the AFRL and DARPA or
the U.S. Government.}}

\IEEEtitleabstractindextext{
\begin{abstract}
In this paper, we explore the capability of both the Adjacency Spectral Embedding (ASE) and the Graph Encoder Embedding (GEE) for capturing an embedded pseudo-clique structure in the random dot product graph setting.
In both theory and experiments, we demonstrate that, in the absence of additional clean (i.e., without the implanted pseudo-clique) network data, this pairing of model and methods can yield worse results than the best existing spectral clique detection methods.
However, these methods can be used to asymptotically localize the pseudo-cliques if additional clean, independent network data is provided.
This demonstrates at once the methods' potential ability/inability to capture modestly sized pseudo-cliques and the methods' robustness to the model contamination giving rise to the pseudo-clique structure.
To further enrich our analysis, we also consider the Variational Graph Auto-Encoder (VGAE) model in our simulation and real data experiments. 
\end{abstract}
}

\maketitle
\IEEEdisplaynontitleabstractindextext
\IEEEpeerreviewmaketitle

\section{Introduction}
Research in the field of graph-structured data is experiencing significant growth due to the ability of graphs to effectively represent complex real-world data and capture intricate relationships among entities. 
Indeed, across various domains including social network analysis, brain tumor analysis, text corpora analysis (e.g., clustering multilingual Wikipedia networks), and citation network analysis, graphs have proven to be a versatile and expressive tool for representing data \cite{girvan2002community,haq2022dacbt,liu2022social,mclaren2022citation}. 
Within the broad field of network analysis, a classical inference task is that of detecting a planted clique/dense subgraph \cite{arora2009computational} in a larger background network.
There is a vast theoretical and applied literature on detecting planted cliques in networks (see, for example, \cite{clique1,clique2,clique3,clique4,feldman2017statistical}), and there are numerous proposed approaches for tackling this problem across a wide variety of problem settings.

In this paper, we explore the planted (pseudo) clique detection problem within the context of spectral methods applied in the setting of random dot product graphs (see Def. \ref{def:rdpg}).
The Random Dot Product Graph (RDPG) \cite{young2007random,athreya2017statistical} is a well-studied network model in the class of latent-position random graphs \cite{hoff_raftery_handcock}.
It is of particular note here due to the ease in which pseudo-clique structures can be embedded into the RDPG model via augmenting the latent positions of the graph with extra (structured) signal dimensions; see Section \ref{sec:pseudo} for detail. 
Estimation and inference in the RDPG framework often proceeds via first embedding the network into a suitably low-dimensional space to estimate the underlying latent structure of the network.

Herein, we consider a trio of graph embedding procedures applied to an RDPG model---the Adjacency Spectral Embedding \cite{STFP-2011}, the Graph Encoder Embedding \cite{shen2022one}, and the Variational Graph AutoEncoder \cite{kipf2016variational}--- with the ultimate goal of better understanding how these methods capture planted clique-like structures in the RDPG model.
The Adjacency Spectral Embedding (ASE) has proven to be a potent modeling/estimation tool for teasing out low-rank structure in complex network data \cite{athreya2017statistical}, with numerous applications and extensions in the literature.
ASE computes a low-rank eigen-decomposition of the adjacency matrix to estimate the latent structure of the RDPG, with this estimated latent structure then forming the foundation on which subsequent inference tasks can be pursued.
The Graph Encoder Embedding (GEE), on the other hand, leverages vertex class labels and edge weight information to specify the vertex embedding via a combination of the adjacency matrix and the column-normalized one-hot encoding of the label vector.
Like ASE, GEE has proven to be an effective tool across a host of inference tasks, e.g., vertex classification, vertex clustering, cluster size estimation, etc. \cite{shen2022one,shen2023graph}.
ASE and GEE can both be seen as spectral embedding tools, and our main results give threshold values for the planted pseudo-clique size under which these methods produce embeddings for the graph with/without the planted pseudo-clique that are effectively indistinguishable,
as well as lower bound thresholds for how the clique signal can be localized in the presence of additional clean, independent network data.
Notably, the indistinguishability thresholds 
are larger than the best known threshold for spectral detection of a planted clique \cite{alon1998finding,manurangsi2020strongish}.

This pairing of model and methods is (to the best of our knowledge) novel, and provides insights into spectral methods useful for both practitioners and researchers (for example, these results could be cast as providing robustness conditions for ASE and GEE under a particular latent position contamination model).
Moreover, recognizing the growing importance of modern graph neural networks, we also incorporate a graph convolutional network-based unsupervised learning approach known as Variational Graph Auto-Encoders (VGAE)\cite{kipf2016variational} into all our experiments to further enrich our findings. 
All relevant Python and R code are available on GitHub \footnote{\url{https://github.com/tong-qii/Clique_detection}}.

\vspace{3mm}
\noindent{\bf Notation:}  For an integer $n>0$, we will use $[n]$ to denote the set $\{1,2,3,\ldots,n\}$.  For a matrix $\mathbf{M}\in\mathbb{R}^{n\times n}$, we will use $\mathbf{M}_{i,\bullet}$ to denote the $i$-th row of $\mathbf{M}$.
We will also routinely use the following matrix norms: the Frobenius norm $\|\mathbf{M}\|_F=\sqrt{\sum_{ij}\mathbf{M}_{ij}^2}$; 
the spectral norm $\|\mathbf{M}\|=\sqrt{\lambda_1(\mathbf{M}^\top \mathbf{M})}$ where $\lambda_1$ denotes the largest eigenvalue of the matrix; and the $2\mapsto \infty$ norm popularized in \cite{cape2toinfty},
$\|\mathbf{M}\|_{2\mapsto\infty}=\sup\{\|\mathbf{M}x\|_{\infty }\text{ s.t. }\|x\|_{2 }=1\}=\max_i \|\mathbf{M}_i\|_2.$
We make extensive use of standard asymptotic notation:.
For two functions $f, g: \mathbb{Z}^+\rightarrow \mathbb{R}^+$, we write $f(n)=o(g(n))$ or $f\ll g$, if $\lim_{n\rightarrow\infty} f(n)/g(n)=0$;
we write $f(n)=\omega(g(n))$ or $f\gg g$ if $g(n)= o(f(n))$; 
we write $f(n)=O(g(n))$ or $f\lesssim g$ if $\exists C>0$ and $n_0$ such that $\forall n\geq n_0$, $f(n)\leq Cg(n)$;
we write $f(n)=\Omega(g(n))$ or $f\gtrsim g$ if $g(n) = O(f(n))$;
we write $f(n)=\Theta(g(n))$ if $f(n)=O(g(n))$ and $g(n)=O(f(n))$.

\section{Pseudo-clique detection in RDPGs}
\label{sec:pseudo}

In this section, we analyze the statistical performance of ASE and GEE to identify an injected (pseudo)clique utilizing the Random Dot Product Graph (RDPG) model defined as follows. 
\begin{defn} 
\label{def:rdpg}
Random Dot Product Graph \cite{young2007random}. Let $\textbf{X} = (X_1, ...,X_n)^T \in \mathbb{R}^{n \times d}$
be a matrix such that the inner product of any two rows satisfies $0 \le X_u^TX_v \le 1$. We say that a random adjacency matrix $\textbf{A}$ is distributed as a random dot product graph with latent positions $\textbf{X}$, and write $\textbf{A} \sim RDPG(\textbf{X})$, if the conditional distribution of $\textbf{A}$ given $\textbf{X}$ is 
\begin{equation}
    \mathbb{P}\hspace{2pt}[\hspace{2pt}\textbf{A} \hspace{2pt}|\hspace{2pt} \textbf{X} \hspace{2pt}] = \prod_{u>v} (X_u^T\hspace{2pt}X_v)^{\textbf{A}_{uv}}\hspace{2pt}(1-X_u^T\hspace{2pt}X_v)^{1-\textbf{A}_{uv}}
\end{equation}
\end{defn}

\noindent RDPGs are a well-studied class of random graph models in the network inference literature, and a large literature is devoted to estimating the RDPG parameters, $\bX$, in the service of subsequent inference tasks such as clustering \cite{lyzinski13:_perfec,STFP-2011,rohe2011spectral}, classification \cite{tangs.:_univer}, testing \cite{tang14:_nonpar,tang14:_semipar,du2023hypothesis}, and information retrieval \cite{zheng2022vertex}, to name a few; see the survey paper \cite{athreya2017statistical} for more applications and exposition of the model in the literature.
We note here that all of our subsequent results follow with minor modifications in the generalized random dot product graph setting of \cite{rubin-delanchy_tang_priebe_grdpg} as well, though this is not pursued further here.
We opt here for the conceptually simpler (and yet still illustrative) RDPG model in our theory, simulations and experiments.

The edge probability matrix of our RDPG is $\bP = \bX \bX^\top$. 
We next augment the latent position $\bX$ matrix with an additional column $\bVc$ to maximally increase the probability of connections between vertices 
in a set $\mathcal{C}$, so that the set of vertices in $\mathcal{C}$ form a denser pseudo-clique in the augmented graph as opposed to their behavior in the original $\bA$.
Note this idea of a rank-1 perturbation introducing a clique/planted structure is not novel, and similar ideas were explored before in various settings; see, for example, \cite{miller2015spectral,montanari2015limitation}.
Formally, 
$$\bVc_i=\begin{cases}
0&\text{ if }i\notin\mathcal{C}\\
\sqrt{1-\sum_{j=1}^d \bX_{i,j}^2}&\text{ if }i\in\mathcal{C}
\end{cases}
$$
The augmented edge probability matrix then becomes $$\bPc = \bXc (\bXc)^\top = \bX \bX^\top + \bVc (\bVc)^\top$$ 
In the sequel, we define
$\alpha(n):=|\mathcal{C}|=\|\bVc\|_0.$
\begin{rem}
Note that if the underlying graph  $\bA$ is a positive semidefinite stochastic blockmodel \cite{holland,karrer2011stochastic} (realized here by having the latent position matrix $\bX$ have exactly $K$ distinct rows), then as long as the vertices in $\mathcal{C}$ are in the same block (i.e., have the same latent position), the above augmentation will a.s. yield a true clique between the vertices in $\mathcal{C}$.
\end{rem}

\subsection{Pseudo-clique detection via Adjacency Spectral Embedding}
\label{sec:ASE}

One of the most popular methods for estimating the latent positions/parameters of an RDPG is the Adjacency spectral embedding (ASE).
The utility of the ASE estimator is anchored in the fact that the classical statistical estimation properties of consistency \cite{athreya2017statistical,lyzinski15_HSBM} and asymptotic normality \cite{athreya2013limit,tang_lse} have been established for the ASE estimator.
\begin{defn}{Adjacency spectral embedding (ASE)}
\label{def:ase}
\cite{athreya_survey,STFP-2011}.
    Given a positive integer $d \geq 1$, the adjacency spectral embedding (ASE) of an adjacency matrix $\bA$ into $\RR^d$ is given by $\hbX = \bU_{\bA} \bS_{\bA}^{1/2}$ where 
    ($\oplus$ denoting the matrix direct sum) 
    \begin{equation*}
    | \bA | = [\bU_{\bA} | \bU_{\bA}^\perp][\mathbf{S}_{\bA} \oplus \mathbf{S}_{\bA}^\perp][\bU_{\bA} | \bU_{\bA}^\perp]^\top
\end{equation*}
is the spectral decomposition of $| \bA | =(\bA ^\top \bA )^{1/2}$ and $\mathbf{S}_{\bA}$ is the diagonal matrix containing the $d$ largest eigenvalues of $| \bA |$ and $\bU_{\bA} $ is the $n \times d$ matrix whose columns are the corresponding eigenvectors. 
We write then that $\hbX=\text{ASE}(\bA,d)$
\end{defn}

The ASE has proven to be a valuable piece of many inferential pipelines including graph matching \cite{patsolic2014seeded,zhang2018unseeded,lyzinski2015spectral}, vertex classification \cite{lyzinski15_HSBM,tangs.:_univer} and anomaly/change-point detection \cite{chen2020multiple,hwang2022bethe,chatterjee2022concentration}, to name a few.
In the context of clique detection---in the Erd\H os-R\'enyi(n,$1/2$) model---spectral methods have proven to be effective for detecting planted cliques down to the (hypothesized) hardness-threshold of clique size $\Omega(n^{1/2})$ \cite{alon1998finding,manurangsi2020strongish}.
Our main theoretical results in the ASE setting (with analogues for GEE in Section \ref{sec:GEE})
demonstrate at once both the difficulty of $\hbXc=$ASE$(\bAc,d)$ in capturing the planted pseudo-clique structure in $\mathcal{C}$ at this threshold level (Theorem \ref{thm:1}) and the capacity of ASE to recover the pseudo-clique when additional latent position information is provided (Theorem \ref{thm:lwb}).  
The theory for ASE below (proven in Appendices \ref{sec:pf1}--\ref{sec:pflwb}) will adapt the $\|\cdot\|_{2\mapsto\infty}$ consistency bound of \cite{athreya2017statistical} for the residual errors when ASE is used to estimate $\bX$.  
Note that the rotations $\bW_\cdot$ appearing in Theorems \ref{thm:1}--\ref{thm:lwb} are a necessary consequence of the rotational nonidentifiability inherent to the RDPG model; i.e., if $\bA\sim$RDPG$(\bX)$ and $\mathbf{B}\sim$RDPG$(\bX\bW)$ for orthogonal matrix $\bW$, then $\bA\stackrel{d}{=}\mathbf{B}$ are equal in law.
\begin{theorem}
\label{thm:1}
Let $\left(\bA_n \sim\text{RDPG}(\bX_n)\right)_{n=2}^\infty$be a sequence of random dot product graphs with $\bA_n$ being the $n \times n$ adjacency matrix. 
Assume that
\begin{itemize}
\item[i.] $\bX_n$ is of rank $d$ for all $n$ sufficiently large;
\item[ii.] There exists constant $a > 0$ such that for all $n$ sufficiently large,
$$
    \delta(\bP_n) := \max_{i}\sum_{j=1}^{n}(\bP_n)_{ij} \geq \log^{4+a}(n)
$$
\item[iii.] There exists constant $c_0 > 0$ such that for all $n$ sufficiently large,
\begin{equation*}
    \gamma(\bP_n) := \frac{\lambda_d (\bP_n)}{\delta(\bP_n)} \geq c_0
\end{equation*}
where $\lambda_d (\bP_n)$ is the $d$-th largest eigenvalue of $\bP_n$;
\item[iv.] (Delocalization assumption) There exists constants $c_1,c_2 > 0$ and a sequence of orthogonal matrices $\widetilde\bW_n$ such that for all $i,j$ and $n$ sufficiently large,
\begin{equation*}
\frac{c_1}{\sqrt{n}}\leq |(\bU_{\bP_n}\widetilde\bW_n)_{i,j}|\leq \frac{c_2}{\sqrt{n}}
\end{equation*}
\end{itemize}
With $\bAc$ the augmentation of $\bA$ as above with $\alpha(n)=o(\delta^{5/6}(\bP_n))$,
let $\hbXc=\text{ASE}(\bAc,d)$ and $\widehat{\bX}=\text{ASE}(\bA,d)$. 
For all $n$ sufficiently large, it holds with probability at least $1-n^{-2}$ that there exists orthogonal transformations $\bW_{n,1},\bW_{n,2} \in \RR^{d\times d} $ such that 
\begin{align*}
    \|&\hbXc\bW_{n,1}- \widehat{\bX}_n\bW_{n,2}\|_{2\mapsto\infty} \\ 
&=O\left(
\frac{\alpha(n)^3  }{ \delta(\bP_n)^{5/2} }+
\frac{\alpha(n)^2 }{\delta(\bP_n)^2}+
\frac{\log n}{\delta(\bP_n)^{1/2}}+
\frac{\alpha(n)}{\sqrt{n \delta(\bP_n) }}
\right)
\end{align*}

\end{theorem}

While it is suspected that detecting planted cliques is hard for the clique size of $o(n^{1/2-\epsilon})$ \cite{manurangsi2020strongish}, this result shows 
an inherent difficulty in detecting a planted pseudo-clique in this pairing of RDPG and ASE 
at the threshold order of $\Omega(n^{1/2})$ if the perturbed graph is embedded into $\mathbb{R}^d$.
Indeed, the above result shows that, for all practical purposes, the embedding of $\bAc$ obtained by ASE into too low of a dimension (here $d$ rather than $d+1$) will be indistinguishable from the embedding $\hbX=ASE(\bA,d)$.
In the dense case where $\|\widehat X_{i,\bullet}\|=\Theta(1)$ for all $i$, the relative signal contained in the clique structure is negligent compared to the signal in the unperturbed latent structure,  
and the pseudo-clique's signal is not well captured in these estimated top $d$ latent space dimensions.

With 
$\alpha(n)=o(\delta^{5/6}(\bP))$, 
elbow analysis/thresholding of the scree plot (adapting the work of \cite{zhu06:_autom} and \cite{chatterjee2014matrix}) for dimension selection in ASE may yield an estimated embedding dimension for $\bAc$ of $d$---rather than the true model dimension of $d+1$---as the $(d+1)$st dimension may appear as a second elbow in the SCREE plot and could be cut by more draconian dimension selection.
In the dense case with $\alpha(n)>>n^{1/2}$, USVT from \cite{chatterjee2014matrix} would likely detect this noise dimension as the USVT threshold is order $\sqrt{n}$, but even this analysis may be complicated in more nuanced data regimes.
For example, in settings where the pseudo-clique is emerging temporally (e.g., $G_1,\cdots,G_m\sim$ RDPG$(\bX$), $G_{m+1}\sim$ RDPG$(\bXc$)), care is needed for embedding the graph sequence into a suitable dimension as choosing the dimension of the embedding of $G_{m+1}$ as that estimated for $G_1$ will effectively mask the pseudo-clique's signal.  

Another way to view the result of Theorem \ref{thm:1} is that ASE is robust to the type of spectral noise introduced by the additional latent space dimension.
In high-noise setting (when we view $\bVc$ as a structured noise dimension), low-rank spectral smoothing via ASE \cite{tang2018connectome} can still capture the signal with high fidelity.

\begin{rem}
An analogous result to Theorem \ref{thm:1} will hold if we augment $\bX$ with $\eta(n)$ disjoint pseudo-cliques each of size at most $\alpha(n)$, and with each pseudo-clique corresponding to the addendum of another additional latent space dimension to $\bX$.  The analogue would then hold (i.e., that the $2\mapsto\infty$ norm of the residuals would $o(1)$ as long as $\eta(n)\alpha(n)$ grows sufficiently slowly; i.e., of order 
$o(\delta^{5/6}(\bP_n))$. 
\end{rem}

\begin{rem}
The result of Theorem \ref{thm:1} may seem to follow immediately from a $2\mapsto\infty$ bounds on 
$\tbXc=\text{ASE}(\bAc,d+1)$, as this could yield concentration of the form
$$
\|\tbXc - \bXc\bW \|_{2\mapsto\infty} =O\left(\frac{\log^2 n}{\delta^{1/2}(\bPc)}\right).
$$
However, the small (as we assume $\alpha(n)=o(\delta^{3/4}(\bP_n))$) eigengap provided by the $(d+1)$st eigenvalue of $\bPc$ complicates direct concentration of the full eigenspace of $\bAc$ to that of $\bPc$.  
Moreover, even if such a result held, the other issue would be that the $d\times d$ principle submatrix of $\bW$ (acting on $\hbXc$) need not be orthogonal as required in Theorem \ref{thm:1}.
  As such, we provide the proof of Theorem \ref{thm:1} in Appendix \ref{sec:pf1}, and note here that it is a relatively straightforward adaptation of the analogous $2\mapsto\infty$ concentration proof in \cite{athreya2017statistical}.
\end{rem}

A partial converse to Theorem \ref{thm:1} is provided by the following result.  First, we establish the following helpful notion of pseudo-clique alignment.
\begin{defn}
\label{def:aligned}
With notation as above, consider $\bA\sim$RDPG$(\bX)$ and a pseudo-clique implanted into $\bA$ via the vector $\bV^{(c)}$.
We say that the pseudo-clique is aligned with $\bP=\bX\bX^T$ if there exists a sequence of orthogonal $(\check\bW_n)$ such that for all $n$ sufficiently large (suppressing the implicit dependence on $n$ below)
\begin{itemize}
\item[i.] For each $j\in [d]$, we have that $\text{sign}( (\bU_\bP\check \bW)_{\ell,j} )=\text{sign}((\bU_\bP\check \bW)_{k,j}  )\text{ for all }\ell,k\in\mathcal{C}$
\item[ii.] For each $i\in [d],j,\in\mathcal{C}$, we have that $|(\bU\bP\check W)_{ij}|=\Theta(n^{-1/2})$
\end{itemize}
\end{defn}
\noindent We then have the following partial converse to Theorem \ref{thm:1}, proven in Appendix \ref{sec:pflwb}.
\begin{theorem}
\label{thm:lwb}
With notation and Assumptions i.--iv. as in Theorem \ref{thm:1}, further assume that $\delta(\bP)=\Theta(n)$, $\alpha(n)=\omega(n^{1/2}\log n)$ and $\alpha(n)=o(n^{3/4})$.
We then have that if the nonzero entries of $\bV$ are $\Theta(1)$ and $\bV$ is aligned with $\bP$, then
there exists sequences of orthogonal transformations $(\bW_{n,1},\bW_{n,2} )$ such that \whp
\begin{align*}
\min_{i\in \mathcal{C}}\|[\hbXc\bW_{n}- \widehat\bX\bW_{n,2}]_{i,\bullet}\|=\Theta(\alpha(n)/n)\\
\max_{i\notin \mathcal{C}}\|[\hbXc\bW_{n}- \widehat\bX\bW_{n,2}]_{i,\bullet}\|=o(\alpha(n)/n).
\end{align*}
\end{theorem}
In the event that the latent positions are a priori known (or a suitably accurate estimate is available, for example via $\widehat\bX$ from an identically distributed graph sample), Theorem \ref{thm:lwb} provides that the pseudo-clique vertices can be asymptotically perfectly recovered by simply ordering the residual errors $\hbXc\bW_{n}- \widehat\bX\bW_{n,2}$.
This contrasts with the result of Theorem \ref{thm:1} which provides a result towards the difficulty of detecting the clique vertices in the absence of this additional information.
We lastly note here that the growth rate in Theorem \ref{thm:lwb} $(n^{3/4})$ is smaller than the general growth rate achieved in Theorem \ref{thm:1}; this is due to the fact that in the setting where $\alpha(n)=\omega(n^{3/4})$, it is markedly more difficult to lower bound the lead order term in the bound of Theorem \ref{thm:1}.  See the proof of Theorem \ref{thm:lwb} for detail. 

\subsection{Detection via the GEE algorithm}
\label{sec:GEE}

Our next results seek to understand how well-competing methods, here the Graph Encoder Embedding (GEE) theoretically and VGAE later empirically, are able to 
preserve the signal contained in an 
implanted pseudo-clique.  While we do not have a corresponding theory yet for VGAE (see Section \ref{sec:exp} for extensive simulations using VGAE), below we derive the analogous theory for GEE.
\begin{defn}{Graph Encoder Embedding (GEE)} \cite{shen2022one}.
   Consider a graph $\bA$ and a corresponding $K$-class label vector for the vertices of $\bA$, denoted $\bY \in \{0, \dots, K\}^n$.
    The number of vertices in each class is denoted by $n_k$, and $\bW \in \RR^{n \times K}$ represents the transformation matrix where $\bW$ is the one-hot encoding of the label vector column-normalized by the appropriate $n_k$. 
    The graph encoder embedding is denoted by $\bZ \in \RR^{n \times K}$ where
    \begin{equation*}
        \bZ = \bA\bW
    \end{equation*}
    and its $i$-th row $\bZ_i$ is the embedding of the $i$-th vertex.  We will write $\bZ=$GEE$(\mathbf{A},\bY)$.
\end{defn}
\noindent In order to make use of the theory for the encoder embedding, we need to consider a slightly modified characterization of the base graph $\bA\sim$RDPG$(\bX)$ from Section \ref{sec:ASE}.
As the GEE requires class labels, we will assume that the rows of the latent positions $\bX$ matrix have an additional class-label feature $\bY$.
This could be achieved, for example, by conditioning on latent positions drawn i.i.d. from a $K$-component mixture distribution, i.e., $(\bX_i,Y_i)\stackrel{i.i.d.}{\sim}F_{XY}$ where $F_{XY}$ is a distribution on $\mathbb{R}^d\times[K]$ and $Y_i$ denotes the mixture component of $\bX_i$; note that our results below consider fixed, not random, latent positions in the RDPG.

Assuming the class labels are identical for $\bA$ and $\bAc$, we have the following result (proven in Appendix \ref{sec:pf2}).
\begin{theorem}
\label{thm:2}
Let $\left(\bA_n \sim\text{RDPG}(\bX_n)\right)_{n=2}^\infty$ be a sequence of random dot product graphs with $\bA_n$ being the $n \times n$ adjacency matrix, and assume the corresponding vertex class vector being provided by $\bY$. 
Assume that $K=\Theta(1)$ and
\begin{itemize}
\item[i.] for each $k\in[K]$, $n_k=\sum_{i=1}^n\mathds{1}\{\bY_i=k\}=\Theta(n)$;
\item[ii.] $\xi(n):=\min_{i,k}\sum_{j=1:\bY_j=k}^{n}(\bP_n)_{ij}=\omega(\sqrt{n\log(n)})
$
\end{itemize}
then with $\bAc$ the augmentation of $\bA$ as above with $\alpha(n)=o(\xi(n))$, let $\bZ=\text{GEE}(\bA,\bY)$ and $\bZc=\text{GEE}(\bAc,\bY)$.  Then with probability at least $1-n^{-2}$ there exists a constant $C>0$ such that
$$
\|\bZc-\bZ\|_{2\mapsto\infty}\leq C\left(\sqrt{\frac{K\log n}{n}}+\frac{\sqrt{K}\alpha(n)}{n}\right).
$$ 
\end{theorem}

Note that under the assumptions of Theorem \ref{thm:2}, a simple application of Hoeffding's inequality yields that the norms of the rows of both GEE embeddings are of order at least 
$\xi(n)/n$ with high probability, and hence the relative error of the difference between the GEE embeddings of $\bAc$ and $\bA$ are $o(1)$.
As in the $\hbXc=$ASE$(\bAc,d)$ setting, in the embedded space GEE is unable to capture the implanted pseudo-clique structure (or clique in the SBM case).
Indeed, the GEEs of $\bA$ and $\bAc$ are asymptotically indistinguishable.
It is of note that the GEE result requires significantly stronger density assumptions than the analogous ASE results, though we suspect this is an artifact of our proof technique.
Weaker bounds (e.g., on $\|\bZc_i-\bZ_i\|_{2}$ for a fixed $i$) are available under much broader sparsity assumptions .

As in the ASE case, we have a partial lower bound for clique structure preservation in the GEE setting (proven in Sec. \ref{sec:pflwb2}), and the pseudo-clique can be localized given (at least an estimate of) the clean/sans pseudo-clique GEE embedding.
\begin{theorem}
\label{thm:lwb2}
With assumptions and notation as in Theorem \ref{thm:2},
assume that $\min_{i,j\in\mathcal{C}}\bP^{(c)}_{ij}=\Theta(1)$, and that $\alpha=\omega(\sqrt{n\log n})$. 
We then have that with high probability
\begin{align*}
\min_{i\in \mathcal{C}}\|[\bZ^{(c)}-\bZ]_{i,\bullet}\|\!=\!\Theta\left(\frac{\alpha}{n}\right),\,\,
\max_{i\notin \mathcal{C}}\|[\bZ^{(c)}-\bZ ]_{i,\bullet}\|\!=\!o\left(\frac{\alpha}{n}\right)
\end{align*}
\end{theorem} 

\begin{figure}[t!]
\begin{center}
    \subfloat[$n_c = \sqrt{n}$]
    {\includegraphics[width=.5\columnwidth]{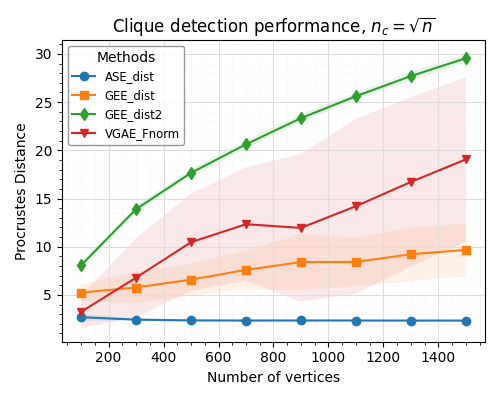}} 
    \subfloat[$n_c = n^{3/4}$]
    {\includegraphics[width=.5\columnwidth]
    {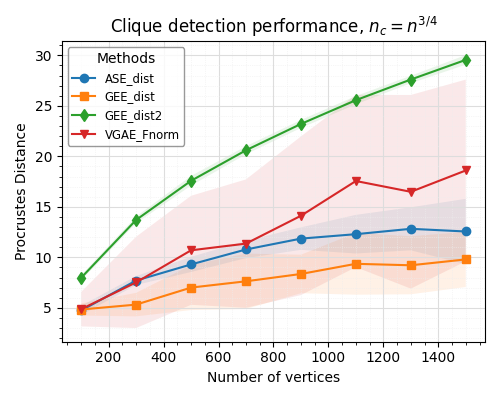}} \\
    \subfloat[$n_c = \sqrt{n} $, normalized]
    {\includegraphics[width=.5\columnwidth]
    {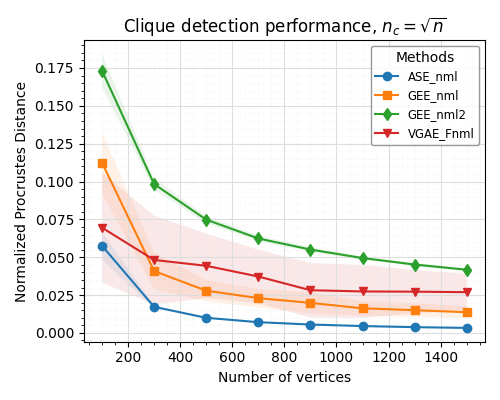}} 
    \subfloat[$n_c = n^{3/4}$,normalized]
    {\includegraphics[width=.5\columnwidth]{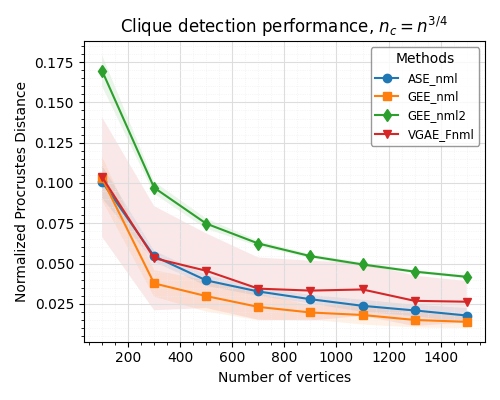}} \\
    \subfloat[Singular values]
    {\includegraphics[width=.5\columnwidth]
    {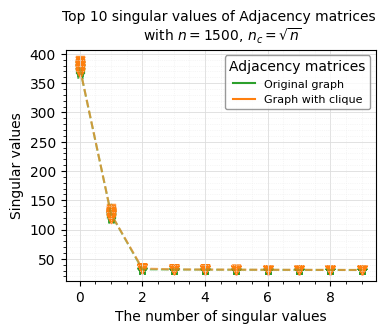}}
    \subfloat[Singular values]
    {\includegraphics[width=.5\columnwidth]
    {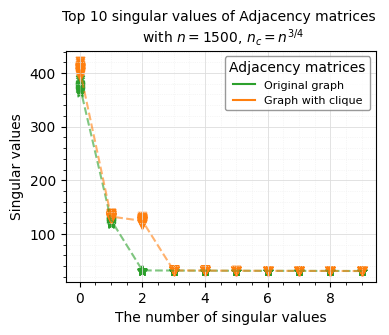}} 
\caption{
The results of unlabeled RDPG design are shown in this figure for the planted pseudo-clique (option ii). The average graph-level distances ($\pm$ 2 s.d.) are displayed in (a) and (b) resulting from $nMC=50$ repetitions, and in (c) and (d) we plot the normalized distances. In (e) and (f), we plot the top 10 singular values of the Adjacency matrices of $\bA$ from $G$ and $\bAc$ from $G^{(c)}$ respectively. The plots for the case where $n_c = \log (n),$ and $n_c=\log^2(n)$ are found in the appendix (Figures \ref{fig:logextra1} and \ref{fig:logextra2}).
  }
  \label{plt:1singular}
\end{center}
\end{figure}

\section{Experiments and simulations}
\label{sec:exp}
In the experiments and simulations to follow, we also incorporate a graph convolutional network-based unsupervised learning approach known as Variational Graph Auto-Encoders (VGAE) \cite{kipf2016variational} into all our experiments to further enrich our findings.
The relationship between ASE and VGAE was recently explored in \cite{priebe2021simple}, and this work motivates our current comparison of the two methods.

Our VGAE framework here, again motivated by \cite{priebe2021simple}, is as follows.
    Consider a graph G=(V,E) with the number of vertices $n = |V|$, adjacency matrix $\bA$, and degree matrix  denoted by $\bD$. 
    The VGAE estimates latent positions $\hbX$ via the variational model
    $p(\hbX|\bX,\bA)=\prod_{i=1}^n p(\hbX_i|\bX,\bA)$
    where (here, we have no latent features so $\bX=I$)
    \begin{align*}
    p(\hbX_i|\bX=I,\bA)&=\text{Norm}(\hbX_i|\mu_i,\text{diag}(\sigma_i^2))
    \end{align*}
   and (here $\tilde{\bA}=\bD^{-1/2}(\bA+I)\bD^{-1/2}$)
    \begin{align*}
    \vec{\mu}&=\text{GCN}_\mu(I,\bA)=\tilde{\bA}\texttt{ReLU}(\tilde{\bA}\bW_0)\bW_1^\mu\\
    \log(\sigma)&=\text{GCN}_\sigma(I,\bA)=\tilde{\bA}\texttt{ReLU}(\tilde{\bA}\bW_0)\bW_1^\sigma
    \end{align*}
    Note that the convolutional networks learn a common first layer of weight and independently learn the second layers.
    To train the VGAE, we aim to optimally reconstruct $\bA$ in the RDPG framework by optimizing the variational lower bound with respect to the weight layers
    $$
\mathbb{E}_{p(\hbX|I,\bA)}\left[ \log \mathbb{P}(\bA|\hbX)-KL[p(\hbX|I,\bA)||q(\hbX)]\right]
    $$
    and where we assume a Gaussian prior $q(\hbX)=\prod_i\text{Norm}(\hbX_i|0,I)$,
    and $KL[q(\cdot)||p(\cdot)]$ is the Kullback-Leibler divergence between $q(\cdot)$ and $p(\cdot)$.
We denote the VGAE embeddings via $\hbX_{VGAE}$ for $\bA$ and $\hbXc_{VGAE}$ for $\bAc$.

\begin{figure}[t!]
\begin{center}
    \subfloat[$n_c =\sqrt{n}$]{\includegraphics[width=.5\columnwidth]{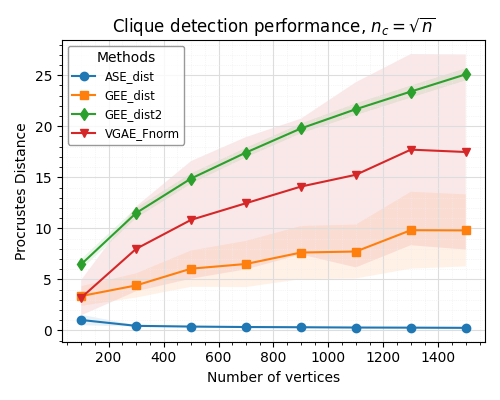}} 
    \subfloat[$n_c = \sqrt{n} $, normalized]{\includegraphics[width=.5\columnwidth]
    {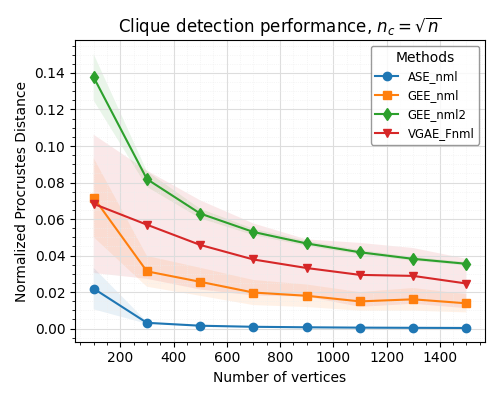}} \\
    \subfloat[$n_c = n^{3/4}$]{\includegraphics[width=.5\columnwidth]{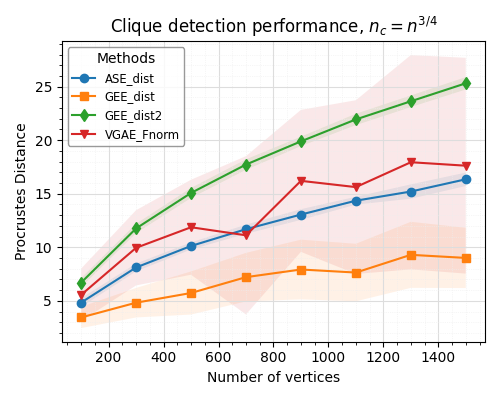}} 
    \subfloat[$n_c = n^{3/4}$, normalized]{\includegraphics[width=.5\columnwidth]
    {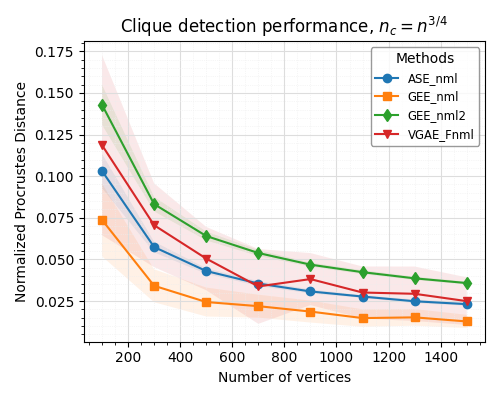}} \\
    \subfloat[$n_c =0.2*n$]{\includegraphics[width=.5\columnwidth]{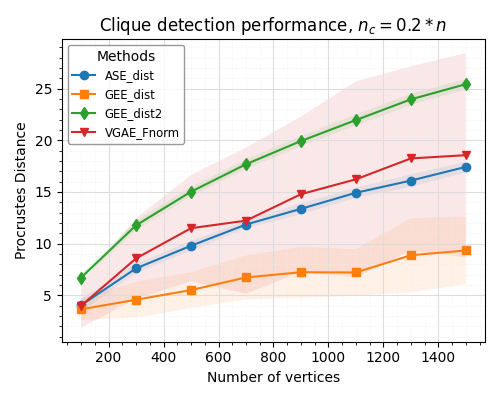}} 
    \subfloat[$n_c = 0.2*n $, normalized]{\includegraphics[width=.5\columnwidth]
    {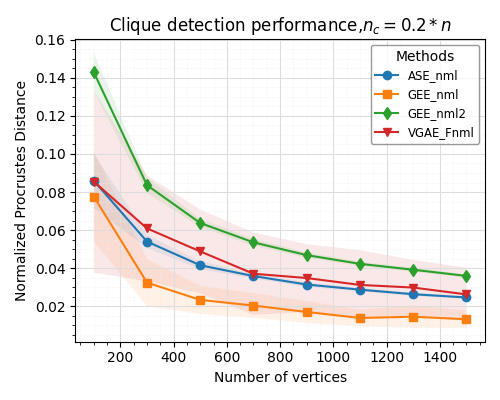}}
\caption{
The results of unlabeled RDPG design are shown in this figure for the planted true clique (option i). The average graph-level distances ($\pm$ 2 s.d.) are displayed in (a) (c) (e) resulting from $nMC=50$ repetitions, and in (b) (d) (f) we plot the normalized distances. The plots for the case where $n_c = \log (n),$ and $n_c=\log^2(n)$ (see Figures \ref{fig:logextra1} and \ref{fig:logextra2}).
  }
\label{plt:1AAc}
\end{center}
\end{figure}

\subsection{Simulations}

We employ simulations to evaluate the strength of the embedded clique-signal when the graph is embedded using three different methods: ASE (Adjacency Spectral Embedding), GEE (Graph Encoder Embedding), and VGAE (Variational Graph Auto-Encoder model). 
To conduct this assessment, we generate pairs of random graphs, labeled as $(G, G^{(c)})$. 
Specifically, $G$ is directly sampled from a specified $n \times 2 $ latent position matrix $\bX$ with i.i.d. rows generated as follows:
\begin{itemize} 
\item[i.] Unlabeled case: Each row is the projection map onto the first two coordinates of an i.i.d. Dirichlet$(1,1,1)$ random variable.
\item[ii.] Labeled case: We consider projecting onto the first two coordinates of i.i.d.\! draws from a 3-component mixture of Dirichlet random variables in $\mathbb{R}^3$.
\end{itemize}
$G^{(c)}$ is derived from $\bXc$ as follows:
\begin{itemize}
    \item[i.] (true clique) A true clique is added between randomly chosen vertices in the RDPG graph in one of the following sizes: $\log(n)$, $\log^2(n)$, $\sqrt{n}$, $n^{3/4}$ or $0.2*n$ vertices; or
    \item[ii.] (pseudo-clique) the additional column $\bVc$  is appended to $\bX$ (yielding an $n \times 3$ matrix) to form a stochastic pseudo-clique in the sampled RDPG (again, we use $\sqrt{1-X_{i1}^2-X_{i2}^2}$ to fill the entries); again this is done in one of these sizes $\log(n)$, $\log^2(n)$, $\sqrt{n}$, $n^{3/4}$ or $0.2*n$ vertices.
\end{itemize}
\begin{figure}[t!]
\begin{center}
  \subfloat[$n_c = \sqrt{n}, K=3$]{\includegraphics[width=.5\columnwidth]{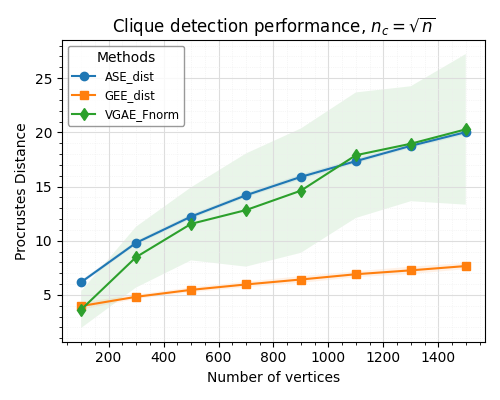}}
    \subfloat[$n_c = \sqrt{n}, K=3 $]{\includegraphics[width=.5\columnwidth]{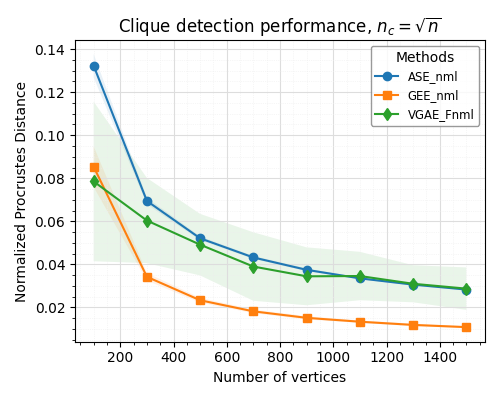}} \\
  \subfloat[$n_c = 0.2*n, K=3$]{\includegraphics[width=.5\columnwidth]{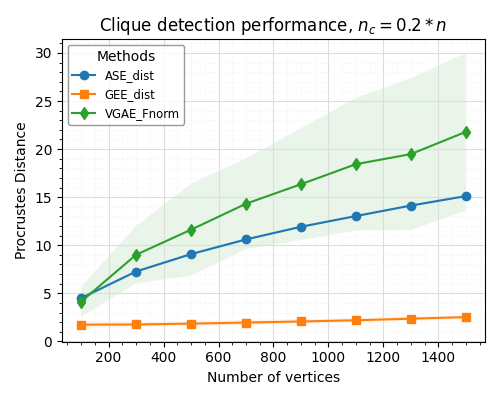}}
    \subfloat[$n_c = 0.2*n, K=3 $]{\includegraphics[width=.5\columnwidth]{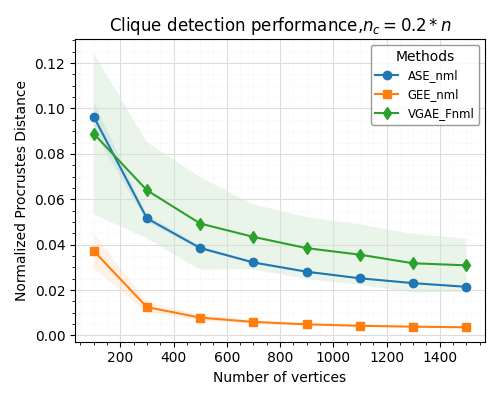}}
\caption{
The results for the clustered RDPG design with $K = 3$ for the planted true clique (option i) are presented in this figure. The experimental configurations remain consistent with those detailed in Figure \ref{plt:1AAc}; i.e., with the planted true clique in the setting where no latent clusters are present.
  }
\label{plt:2K3}
\end{center}
\end{figure}
The total number of vertices in these graphs varies within the range of $n=[$100, 300, 500, $\dots, 1500]$. 
For each pair $(G, G^{(c)})$, we apply ASE, GEE, and VGAE, followed by computing relevant distances between the resulting graph embeddings. 
Specifically, we compute the following.
For ASE we calculate the graph-wise (here  $D_{\text{Proc}}(\hbXc,\hbX)=\min_{\bW\in\mathcal{O}_d}\|\hbXc-\hbX\bW\|_F$) and vertex-wise Procrustes distance (see Eq. \ref{eq:proclocal}) accounting for the nonidentifiability of the ASE estimator.
For the GEE, we compute the Procrustes distances (again at the graph and vertex levels) of the embedding $\min_{\bW\in\mathcal{O}_d}\|\bZc-\bZ\bW\|_F$ to account for the possible differences in cluster labelings across the GEE embeddings. 
For the VGAE, we compute the Frobenius norm distances (again at the graph and vertex levels) of the embedding directly $\|\hbXc_{VGAE}-\hbX_{VGAE}\|_F$.
This entire process is repeated $nMC = 50$ times for each pair of graphs, and the average distances are recorded with error bands representing $\pm2$s.d.

We note here that we are not running a clique-detection algorithm on the planted-clique embedding;  
for a comparison of how many commonly used dense subgraph detection algorithms fare in detecting the pseudo-clique, see Appendix \ref{app:SOTA}.  
We are rather comparing the embeddings (at the graph-level and at the vertex-level) before and after the clique is implanted. 
If the before and after graph embeddings are relatively indistinguishable, then this is evidence that the clique structure is not captured well in the embedding space.
Even in the event that the graph-level embeddings of the before and after graphs are sufficiently different, if the distances across the two embeddings for the clique vertices (versus the non-clique vertices) at the vertex-level are not localized (i.e., large values are not concentrated on the clique vertices) then the signal in the clique may be present, but the clique vertices are difficult to accurately identify.
As we will demonstrate below, different methods (ASE, GEE, and VGAE) exhibit markedly different performances across the experimental conditions.

\begin{figure}[t!]
\begin{center}
    {\includegraphics[width=.49\columnwidth]{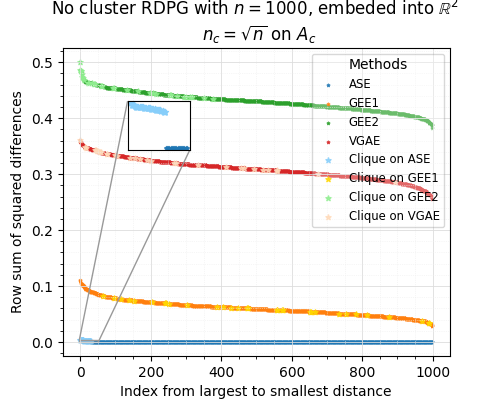}} 
    {\includegraphics[width=.49\columnwidth]
    {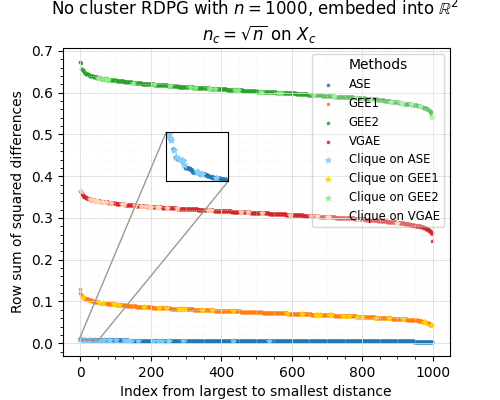}} \\
{\includegraphics[width=.49\columnwidth]
    {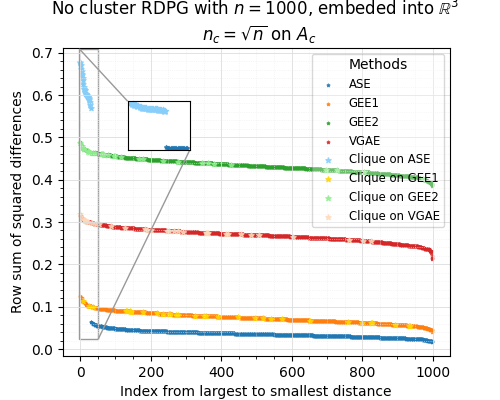}}
    {\includegraphics[width=.49\columnwidth]
    {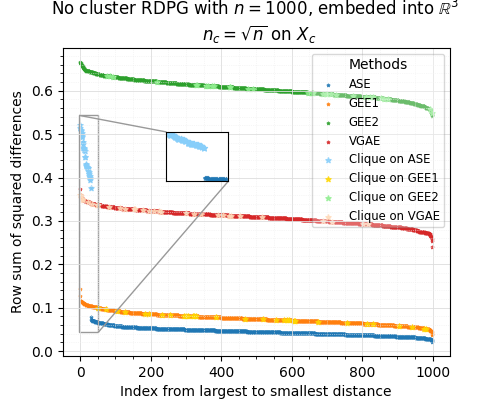}}

\caption{
The results for planted true clique (column 1) and the planted pseudo-clique (column 2) all of size $n^{1/2}$. The \emph{vertex-level} distances are displayed in row 1 (resp. 2) embedding the graphs into $\mathbb{R}^2$ (resp., $\mathbb{R}^3$) for ASE and VGAE. Results are averaged over $nMC=50$.
GEE is embedded into the dimension equal to the number of clusters fed into the algorithm; GEE is included in $\mathbb{R}^2$ and $\mathbb{R}^3$  panels for reference.
}
\label{fig:sqrtn_local}
\end{center}
\end{figure}

\begin{figure}[t!]
\begin{center}
    {\includegraphics[width=.49\columnwidth]
    {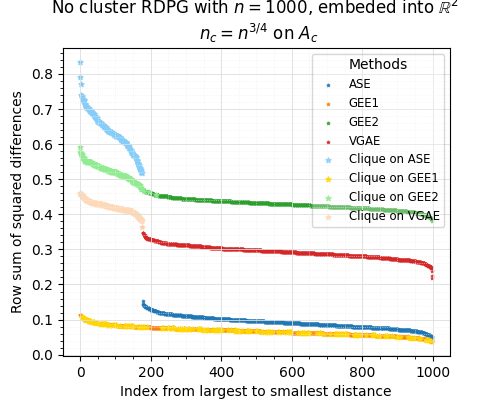}} 
    {\includegraphics[width=.49\columnwidth]{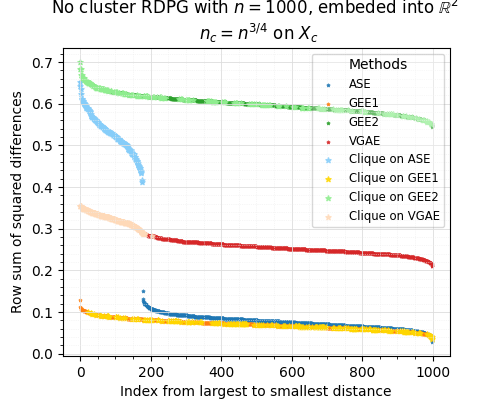}} \\
    {\includegraphics[width=.49\columnwidth]
    {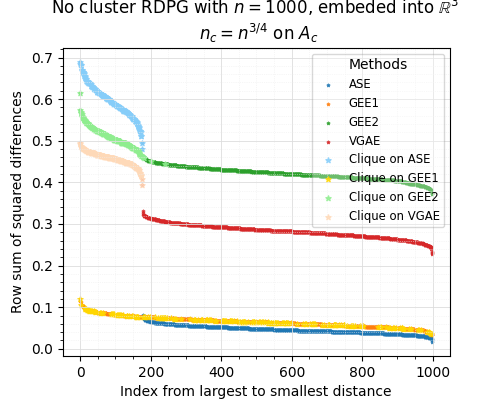}} 
    {\includegraphics[width=.49\columnwidth]
    {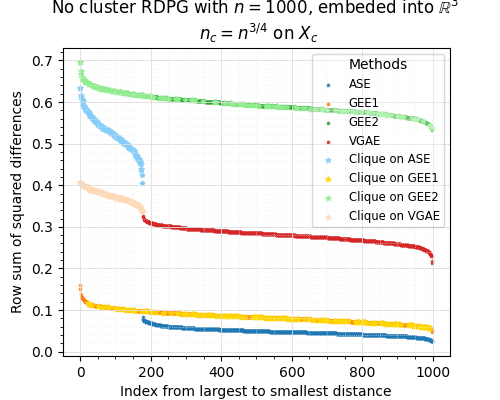}} 
\caption{
The results for planted true clique (column 1) and the planted pseudo-clique (column 2) all of size $n^{3/4}$. The \emph{vertex-level} distances are displayed in row 1 (resp. 2) embedding the graphs into $\mathbb{R}^2$ (resp., $\mathbb{R}^3$) for ASE and VGAE. Results are averaged over $nMC=50$.
GEE is embedded into the dimension equal to the number of clusters fed into the algorithm; GEE is included in $\mathbb{R}^2$ and $\mathbb{R}^3$  panels for reference.
}
\label{fig:34dim23}
\end{center}
\end{figure}

\begin{figure}[t!]
\begin{center}    
    {\includegraphics[width=.49\columnwidth]
    {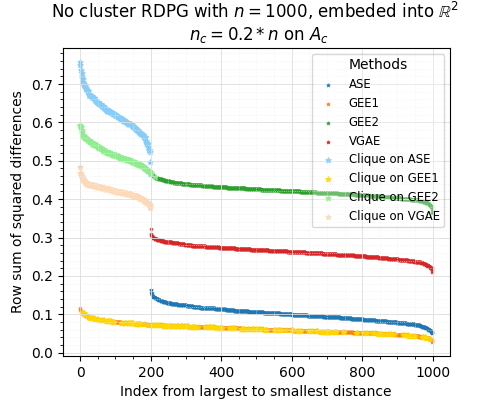}} 
    {\includegraphics[width=.49\columnwidth]{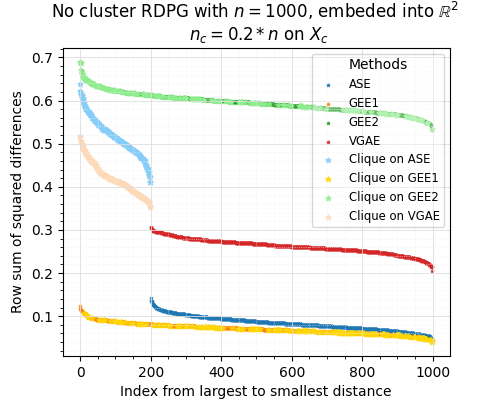}} \\
    {\includegraphics[width=.49\columnwidth]
    {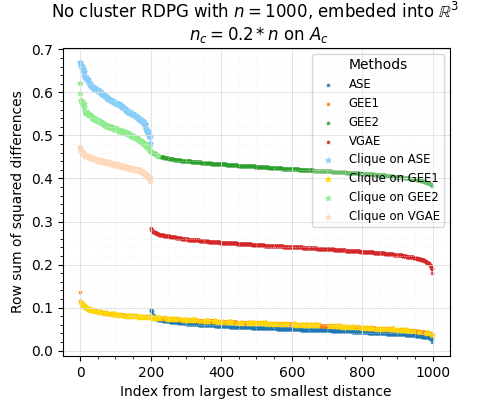}} 
    {\includegraphics[width=.49\columnwidth]
    {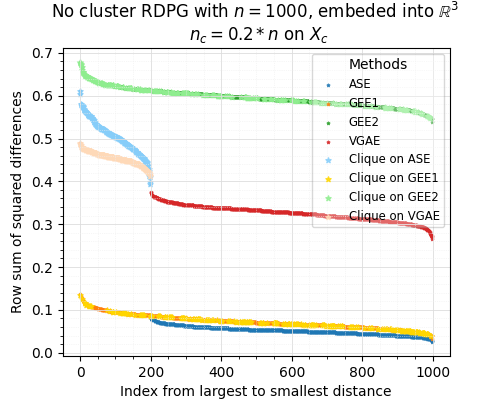}}    
\caption{
The results for planted true clique (column 1) and the planted pseudo-clique (column 2) all of size $0.2\!*\!n$. The \emph{vertex-level} distances are displayed in row 1 (resp. 2) embedding the graphs into $\mathbb{R}^2$ (resp., $\mathbb{R}^3$) for ASE and VGAE. Results are averaged over $nMC=50$.
GEE is embedded into the dimension equal to the number of clusters fed into the algorithm; GEE is included in $\mathbb{R}^2$ and $\mathbb{R}^3$  panels for reference.
}
\label{fig:02dim23}
\end{center}
\end{figure}

We conducted two sets of experiments.
The results of the first experiment are visualized in Figure \ref{plt:1singular} (and Figures \ref{plt:1AAc}, \ref{fig:sqrtn_local}--\ref{fig:02dim23}). 
Therein, we sampled data from an RDPG latent position matrix without any predefined cluster structure. 
Within this set, we employed two variants of the Leiden algorithm \cite{traag2019louvain} to identify node clusters in order to apply GEE.  
Note that $\bA$ and $\bAc$ are clustered separately, and hence the cluster labels/assignments may differ across graphs, and the Procrustes distance is used to ameliorate this.
The first approach, known as the modularity vertex partition employs modularity optimization and often yields a smaller number of clusters. 
Consequently, the distances between graph embeddings are also often smaller, as indicated by the \texttt{GEE\_dist} (or \texttt{GEE1}) values shown in Figure \ref{plt:1singular}. 
The second approach, referred to as the CPM vertex partition, implements the Constant Potts Model and often yields a larger number of clusters.
This results in relatively larger across-graph distances (due to variability in label/cluster assignment and not necessarily the clique signal strength), as represented by the \texttt{GEE\_dist2} (or \texttt{GEE2}) values in Figure \ref{plt:1singular}.
In the second batch of experiments (see, for example, Figure \ref{plt:2K3}) we sampled the RDPG with a fixed number of clusters ($K=3$). In this scenario, as the number of clusters is predetermined (we still use Leiden to identify the clusters), only one GEE distance measurement is recorded. 

In Figure \ref{plt:1singular}, panel (a) demonstrates that ASE struggles to capture the gross pseudo-clique 
perturbation 
when the pseudo-clique size is relatively small, here $\sqrt{n}$. 
That said, in this small pseudo-clique setting, ASE is better able to capture the clique signal (locally and globally) if the graph is embedded into $\mathbb{R}^3$ as opposed to $\mathbb{R}^2$ 
particularly in the planted true clique design  
(see Figure \ref{fig:sqrtn_local}). 
Even in the relatively larger pseudo-clique settings where ASE is better able to differentiate the embeddings pre/post planted pseudo-clique (panel b), the relative strength (distances divided by $\|\hbX\|_F$) of the signal of the pseudo-clique in the embedding is rapidly diminishing (panel d).
As the graph size increases, GEE with a greater number of clusters (\texttt{GEE\_dist2}) consistently shows larger distances on average and better captures the gross pseudo-clique perturbation compared to VGAE and GEE with fewer clusters; however this is at the expense of relatively poor pseudo-clique localization in the embedding as the differences across the embeddings are more spread among all the vertices in the graph (see Figures \ref{fig:sqrtn_local}--\ref{fig:02dim23}). 
VGAE performs well in capturing the gross pseudo-clique signal, but again does not localize the clique signal well in the small pseudo-clique size settings (see Figures \ref{fig:sqrtn_local}).
In the relatively larger clique setting ($\geq n^{3/4}$) ASE and VGAE
appear to better balance capturing gross pseudo-clique structure and localizing the clique signal in the embedding.

\begin{figure}[t!]
\begin{center}
    {\includegraphics[width=.49\columnwidth]{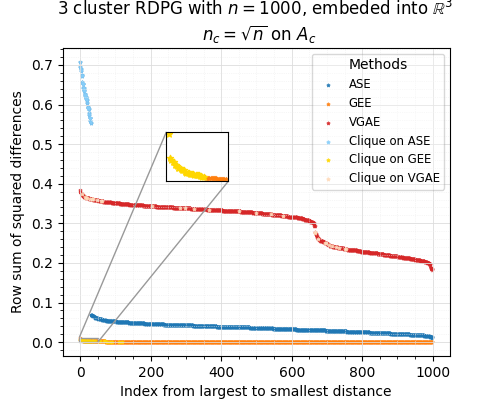}} 
    {\includegraphics[width=.49\columnwidth]
    {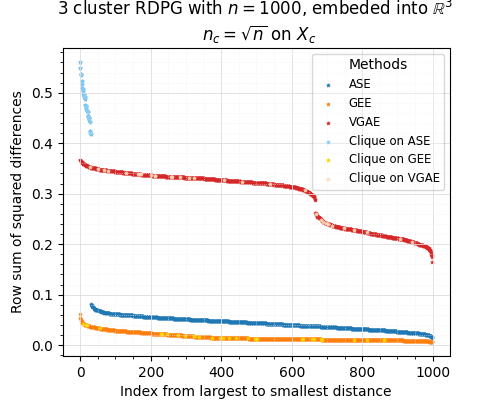}} \\
    {\includegraphics[width=.49\columnwidth]
    {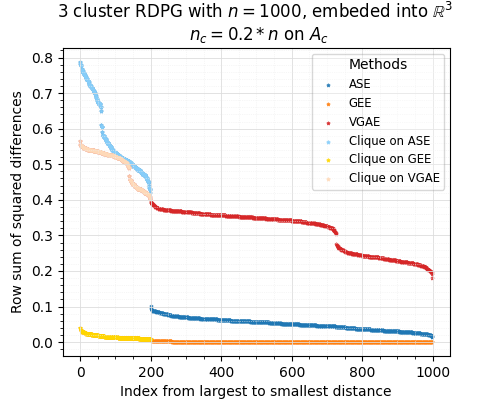}} 
    {\includegraphics[width=.49\columnwidth]{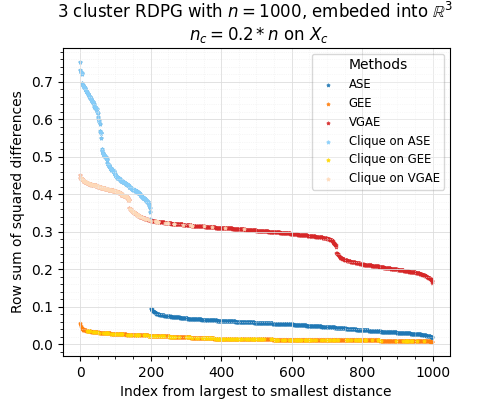}}  
    
\caption{
The results for planted true clique (column 1) and the planted pseudo-clique (column 2) size $\sqrt{n}$ in row 1 and $0.2*n$ in row 2. The \emph{vertex-level} distances are displayed in row 1 (resp. 2) embedding the graphs into $\mathbb{R}^3$ for ASE and VGAE. Results are averaged over $nMC=50$.
GEE is embedded into the dimension equal to the number of clusters fed into the algorithm; GEE is included in $\mathbb{R}^2$ and $\mathbb{R}^3$  panels for reference.
}
\label{plt:5}
\end{center}
\end{figure}

Our findings for the planted true clique are depicted in Figure \ref{plt:1AAc} and align closely with those depicted in Figure \ref{plt:1singular}.
Moreover, in the context of a three-cluster RDPG design ($K=3$), the findings depicted in Figure \ref{plt:2K3} (a) indicate that VGAE and ASE yield the largest distances among the methods; i.e., they capture the gross true-clique perturbation best in the embedding space, and both also localize the clique vertices well (see Figure \ref{plt:5}).
Notably, GEE consistently produces the smallest distances when the number of clusters is fixed (this is similar to the \texttt{GEE1} case), and it demonstrates lower variance when compared to the no-cluster RDPG design. 

In addition to calculating the distances before and after pseudo-clique planting, we also explore the top 10 singular values of the adjacency matrices $\bA$ and $\bAc$. 
For illustration, we focus on graphs with a size of $n=1500$ as an example, and we depict the top 10 singular values across $nMC=50$ simulations in Figure \ref{plt:1singular} (e) and (f). 
The results reiterate that when the clique size is small (here $\sqrt{n}$), the top singular values of $A_1$ and $A_2$ are nearly identical, though
interestingly, in this small clique setting ASE is better able to capture the clique signal if the graphs are embedded into $\mathbb{R}^3$ here (though dimension selection via elbow analysis of the singular value scree plot would likely yield $d=2$).
In contrast, with the introduction of a bigger-sized clique, the top three singular values of $A_2$ surpass those of $A_1$.
In these larger clique settings ASE seems to localize the clique signal equally well in $\mathbb{R}^3$ and in $\mathbb{R}^2$ (see Figures \ref{fig:34dim23}--\ref{fig:02dim23}), indicating the clique's signal has bled into the higher eigen-dimensions.

To examine the findings of Theorems \ref{thm:lwb} and \ref{thm:lwb2}, we consider a  
more detailed examination of the precise distance between the estimated latent positions of individual vertices, we compute the sum of squared distances for each row between the embeddings of $G$ and $G^{(c)}$, which is represented as (where $\bW$ is the Procrustes alignment in the ASE and GEE cases)
\begin{align}
\label{eq:proclocal}
d^{(ASE)}_i &= |\hbX_i - \bW(\hbXc)_i|^2\\
\label{eq:proclocal2}
d^{(VGAE)}_i &= |(\hbX_{VGAE})_i - (\hbXc_{VGAE})_i|^2\\
\label{eq:proclocal3}
d^{(GEE)}_i &= |\bZ_i - \bW\bZc_i|^2
\end{align} 
Results are displayed in Figures \ref{fig:sqrtn_local}--\ref{plt:5}.
In each figure, we plot the performance with the planted true clique (column 1) and the planted pseudo-clique (column 2).  
In Figures \ref{fig:sqrtn_local}--\ref{fig:02dim23}, these \emph{vertex-level} distances are displayed in row 1 embedding the graphs in to $\mathbb{R}^2$ and in row 2 embedding the graphs in to $\mathbb{R}^3$ for ASE and VGAE 
(darker colors denote non-pseudo-clique vertices, lighter colors pseudo-clique vertices)
; note that GEE is embedded into the dimension equal to the number of clusters fed into the algorithm (here deduced via a Leiden clustering), and GEE is included in all panels of these figures for ease-of-reference. 
From Figure \ref{fig:sqrtn_local}, we see that 
ASE seems to do the best job of localizing the signal in the clique structure when the clique is relatively small in size, but the pseudo-clique signal is quite weak particularly when embedded into $\mathbb{R}^2$  
(see the inset plots for a zoom in of the structure on row 1).  
That said, the level of the global signal in the ASE embeddings is weak (relative to other methods) and the noise across the embeddings still greatly inhibits the discovery of these clique vertices in a single embedding.
When the embedding dimension is increased to $\mathbb{R}^3$, the true latent dimension for the pseudo-clique setting, the ASE better localizes the pseudo-clique vertices.
VGAE, \texttt{GEE1} and , \texttt{GEE2} seem to struggle localizing the clique/pseudo-clique structure in the small clique size setting, though these methods do exhibit a large global change in the graph pre/post pseudo-clique implantation.   
This figure indicates the global change may be a result of variance in the method rather than entirely from the clique signal.
Another trend to note is that the performance of VGAE is relatively insensitive to embedding the data into $d=2$ dimensions or the true $d=3$ dimensions.

While ASE and \texttt{GEE2} (in the true clique setting) localize well in the large clique setting ($0.2n$; Figure \ref{fig:02dim23}), VGAE's performance is also of note here. 
VGAE exhibits a threshold at 200 in the true clique setting (faintly visible also in the $n^{3/4}$ setting), where there is a notable jump from the non-clique-vertex distances to the clique-vertex distances.
Of note is that this jump is not as prominent in the model-based planted pseudo-clique setting.
In the $K=3$ class RDPG setting, ASE and VGAE (and not GEE) were able to capture the global structural change caused by the pseudo-clique;
all three did well localizing the true planted clique (see Figure \ref{plt:2K3}), and we see that in the localized distances as well; see Figure \ref{plt:5}.
Again, in the larger true-clique case, VGAE and ASE exhibit an interesting threshold at 200 evincing its ability to balance the preservation of local and global structure. 
That said, in the smaller clique-size case, the clique signal in the VGAE embedding is less localized than in that of ASE (and even GEE in cases).
\begin{table}[t!]
    \centering
    \begin{tabular}{c|c | c |c | c|c|c}
    \hline
    \hline
        $n_c$ & ASE & GEE1 & GEE2 & GAE & Type &d\\
        \hline
        $\log(n)$&0.0120 & 0.0103 & 0.0124 & 0.0175 &Pseudo&$\mathbb{R}^2$ \\
        $\sqrt{n}$ & 0.0463 & 0.0370 & 0.0389 & 0.0626&Pseudo&$\mathbb{R}^2$ \\
        $n^{2/3}$ & 0.6856 & 0.0988 & 0.1245 & 0.3267&Pseudo&$\mathbb{R}^2$\\
        $n^{3/4}$ & 0.8346 & 0.1775 & 0.2330 & 0.5862&Pseudo&$\mathbb{R}^2$ \\
        $0.2n$ & 0.8103 & 0.1943 & 0.2507 & 0.6870&Pseudo&$\mathbb{R}^2$ \\
        \hline 
        $\log(n)$ & 0.0151& 0.0135 & 0.0131& 0.0098&Pseudo&$\mathbb{R}^3$\\
        $\sqrt{n}$ & 0.7313 & 0.0341 & 0.0381 & 0.0461&Pseudo&$\mathbb{R}^3$ \\
        $n^{2/3}$ & 0.9224 & 0.1006 & 0.1236 & 0.2794&Pseudo&$\mathbb{R}^3$\\
        $n^{3/4}$ & 0.9094 & 0.1752 & 0.2239 & 0.5602&Pseudo&$\mathbb{R}^3$ \\
        $0.2n$ & 0.9029 & 0.1930 & 0.2530 & 0.6744&Pseudo&$\mathbb{R}^3$ \\
    \hline
    \hline
    \end{tabular}
    \captionof{table}{
    Mean Average Precision (MAP) for detecting the planted true/pseudo-clique in unlabeled RDPG models in $\mathbb{R}^2$ and $\mathbb{R}^3$ over $nMC = 50$ simulations. Pseudo-clique sizes considered are $n_c = [\log(n), \sqrt{n}, n^{2/3}, n^{3/4}, 0.2n]$.
    Graphs are 1000 vertices.
    }
    \label{table:MAP}
\end{table}

In all settings and methods, we see that the pseudo-clique is more difficult to localize than the true planted clique.
To explore pseudo-clique detectability further, we evaluate performance using the mean average precision (MAP), computed as the average precision of retrieving the true clique or pseudo-clique vertices based on the similarity rankings defined in Eq. \ref{eq:proclocal}--\ref{eq:proclocal3} aggregated over $nMC=50$. 
The MAP is a standard metric in ranking and retrieval tasks; higher values indicate better performance, with \text{MAP} close to 1 corresponding to perfect precision (i.e., all clique or pseudo-clique vertices ranked at the top positions). 
From the table, we see that all methods struggle to localize small pseudo-cliques, with markedly better localization for larger pseudo-cliques.
ASE seems to localize the best here, and we note that 
ASE is more effectively able to capture the planted structure when embedded into $\mathbb{R}^3$ compared to $\mathbb{R}^2$.

\subsection{EUemail}
\label{sec:EU}

\begin{figure}[t!]
    \centering
    \includegraphics[width =0.7\columnwidth]{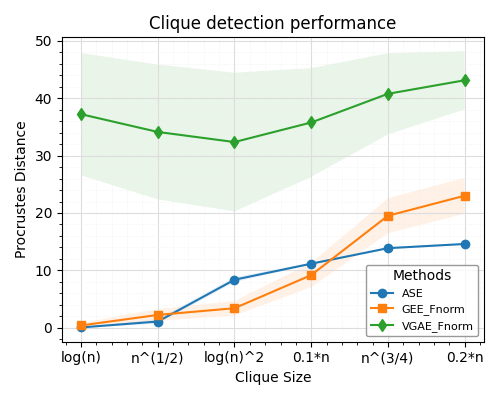}
    \caption{Visualization of the average distances between embeddings with and without clique for EUemail data.}
    \label{fig:eu}
\end{figure}

The \texttt{EUemail} dataset is a network created from email data obtained from a prominent European research institution. It is comprised of $1005$ nodes, representing members of the institution. An edge $\{u,v\}$ exists between two individuals if either person $u$ has sent at least one email to person $v$ or vice versa. Each individual belongs to one of the 42 departments within the research institution. The dataset also includes the ground-truth community membership information for the nodes \cite{yin2017local}; note that in GEE we use these 42 ground-truth communities in both the original and clique-implanted embeddings, and so here the Procrustes step is not needed to ameliorate the different community labels.
Applying a methodology akin to our experimental approach, we randomly select a subset of vertices to form a true clique in the original graph. 
The number of vertices in the clique varies as $n_c = [\log(n), \sqrt{n}, \log(n)^2, 0.1n, n^{3/4}, 0.2n]$. 
Subsequently, we compute the distances between the original graph embedding ($d$ in ASE estimated using elbow analysis of the SCREE plot) and the embedding of the graph containing the clique using ASE (with the same $d$), GEE and VGAE methods. 
The entire process is repeated $nMC = 50$ times with error bands $\pm2$s.d., and we plot the average distances in Figure \ref{fig:eu}. The outcomes are also summarized in Table \ref{table:eu_gae}.

\begin{table}[t!]
    \centering
    \begin{tabular}{c|c | c |c }
    \hline
    \hline
        $n_c$ & ASE & GEE & VGAE \\
        \hline
        $\log(n)$ &0.0434 & 0.3923 & 37.2258   \\
        $\sqrt{n}$ & 1.0743 & 2.2252 & 34.1321 \\
        $\log(n)^2$ &8.3648& 3.3992 & 32.3718 \\
        $0.1n$ & 11.1573 & 9.1703 & 35.7826 \\
        $n^{3/4}$ & 13.8667 & 19.5171 & 40.7796 \\
        $0.2n$ & 14.5977 & 23.0077 & 43.1421 \\
    \hline
    \hline
    \end{tabular}
    \captionof{table}{Using the EUemail graph dataset, we randomly select a set of vertices to form a clique, the size of it varies as $n_c = [\log(n), \sqrt{n}, \log(n)^2, 0.1n, n^{3/4}, 0.2n]$. We compute the distances between the original graph and the graph with clique using methods ASE, GEE and VGAE. This figure shows the record of the mean distance ($\pm$ 2 s.d.) after doing $nMC = 50$ simulations.}
    \label{table:eu_gae}
\end{table}

When the clique size is small, ASE's inability to distinguish between two graph embeddings aligns with both theoretical expectations and empirical observations. As the clique size increases, ASE progressively discerns disparities between the embeddings. Meanwhile, GEE consistently produces increased distances with a fixed 42 clusters. Remarkably, VGAE seems to extract a wealth of information from the complex data within the real graph dataset, resulting in significantly larger distances across embeddings when compared to the other methods. However, it also exhibits a noteworthy degree of variability, particularly when contrasted with the more stable outcomes of GEE and ASE.

\section{Conclusion}
\label{sec:concl}
Graph embedding techniques are widely used to create lower-dimensional representations of graph structures, which can be used for multiple downstream inference tasks such as community detection and graph testing. 
In this paper, we theoretically investigate the capabilities of ASE and GEE for use in capturing the signal contained in pseudo-cliques planted into RDPG latent position graphs. 
Our experimental findings support and augment these theoretical insights: namely that ASE (embedded into too low a dimension) and GEE (with a fixed cluster structure) poorly capture---both locally and globally---the pseudo-clique signal when the pseudo-clique is small.
Both do exhibit better performance capturing the pseudo-clique signal when the clique size grows larger, though
the influence of clique size is not the sole factor at play.
We note that community membership information and the number of communities noticeably affect the performance of GEE. 
A higher degree of clustering variability (often in the case of a larger number of clusters) accentuates the global distinguishability between the graphs $\bA$ and $\bAc$, often at the expense of the distances across embeddings not localizing on the clique vertices. 
The VGAE, while unexplored theoretically, experimentally exhibits its own performance quirks.
The method is best able to detect large planted true cliques, though struggles to balance global signal capture and local signal concentration in the smaller clique and model-based planted clique  settings. 

Beyond deriving the analogous theoretical results to Theorems \ref{thm:1}---\ref{thm:lwb2} in the VGAE setting and for related embedding methods (e.g., the Laplacian Spectral Embedding \cite{rohe2011spectral,tang_priebe_16}), it is natural to also consider planting (and detecting) pseudo-cliques in more general random graph models.  
Moreover, as mentioned previously, the results contained herein could also be cast as a type of robustness of ASE and GEE to a particular noise model.
Furthering these results to understand the robustness/detection ability of ASE and GEE to additional structures that can be embedded into the graph model (e.g., higher degree of transitivity \cite{rohe2013blessing}) is a natural next step.

\bibliographystyle{plain}
\bibliography{bibs_pro}

\vspace{-2cm}
\begin{IEEEbiography}[{\includegraphics[width=1in,keepaspectratio]{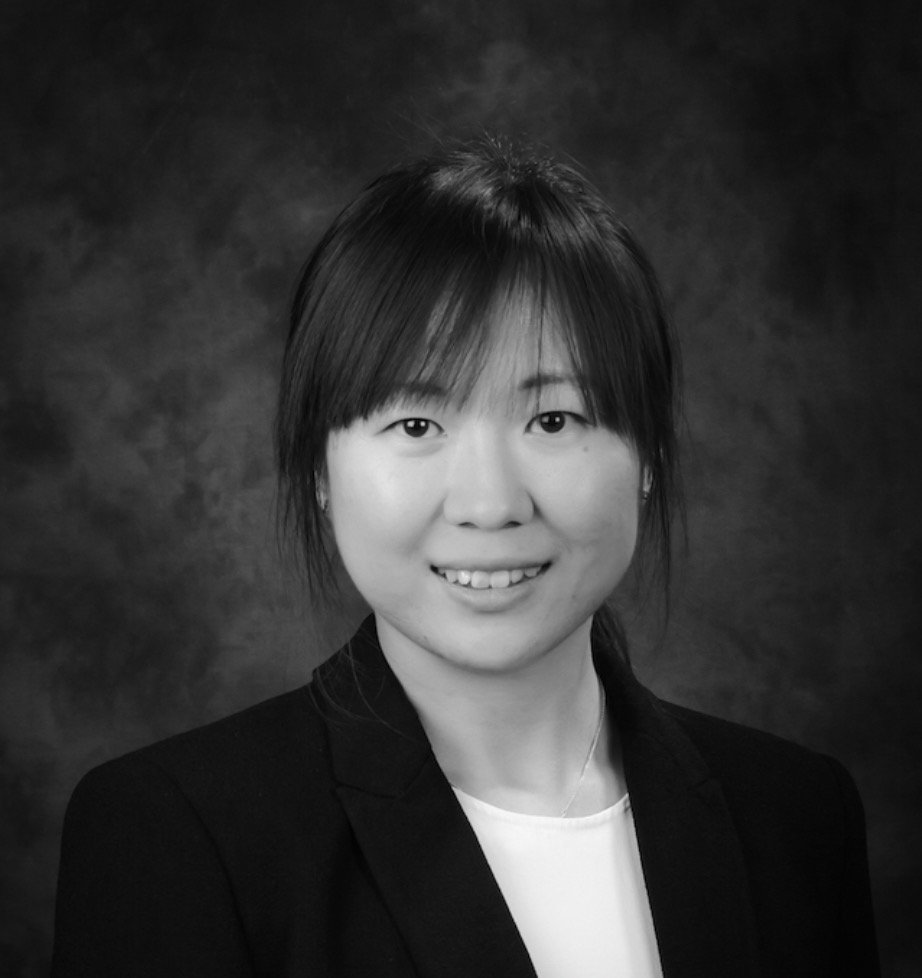}}]{Tong Qi}
 received the BS degrees in agricultural economics from Purdue University in 2013, the MS degree in management from Purdue University in 2015, the MS degree in joint statistics and computer science from Purdue University in 2019, and the Ph.D degree from the STAT program at the University of Maryland, College Park in 2025.  Her research interests include graph embedding, statistical network analysis, and machine learning. Webpage: \url{https://tong-qii.github.io/}.
\end{IEEEbiography}

\vspace{-8mm}
\begin{IEEEbiography}[{\includegraphics[width=1in,height=1.25in,clip,keepaspectratio]{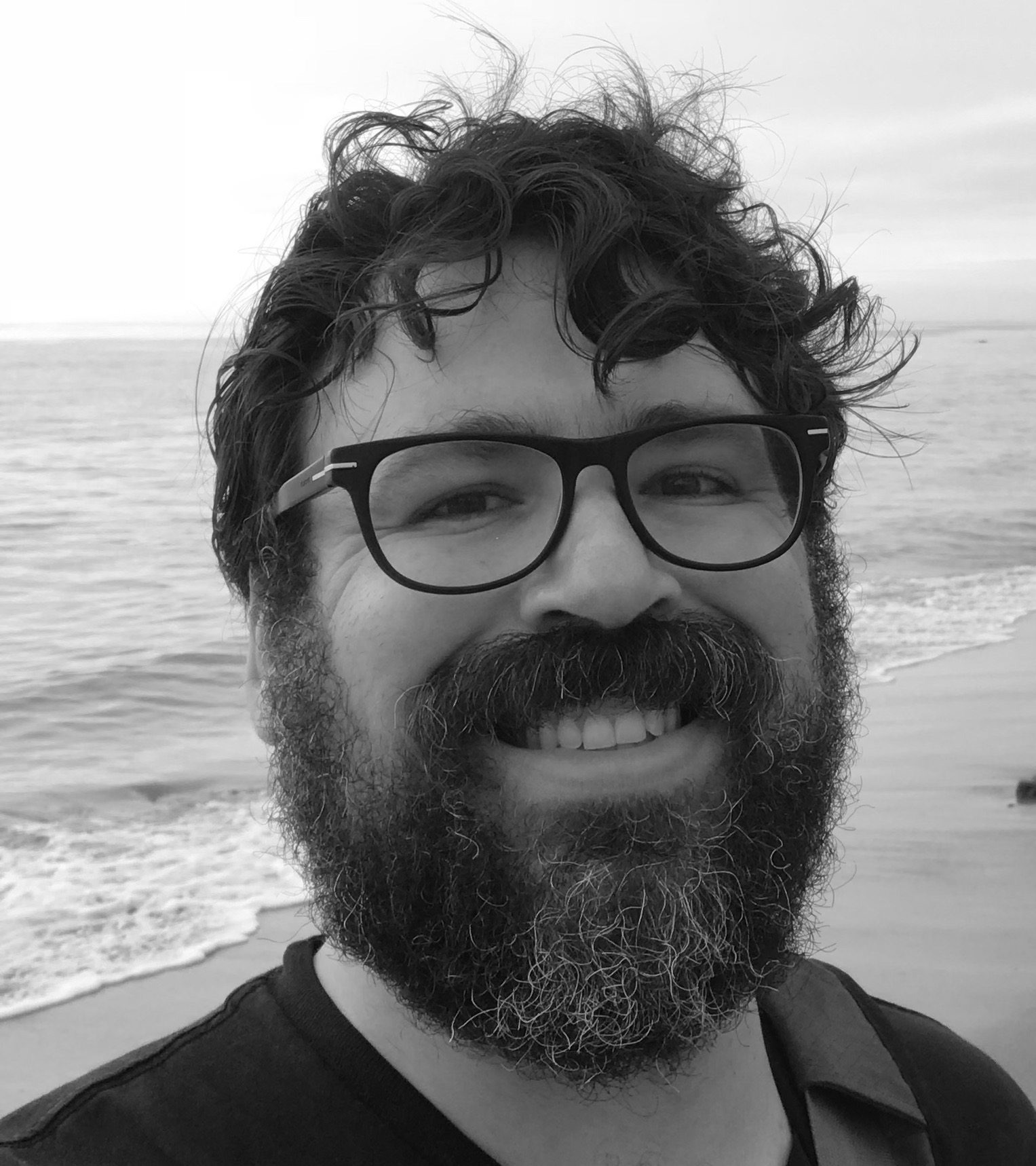}}]{Vince Lyzinski}
 received the BSc degree in mathematics from the University of Notre Dame, in
2006, the MA degree in mathematics from Johns
Hopkins University (JHU), in 2007, the MSE
degree in applied mathematics and statistics
from JHU, in 2010, and the PhD degree in applied
mathematics and statistics from JHU, in 2013.
From 2013-2014 he was a postdoctoral fellow
with the Applied Mathematics and Statistics
(AMS) Department, JHU. During 2014-2017, he
was a senior research scientist with the JHU
HLTCOE and an assistant research professor with the AMS Department, JHU.
From 2017-2019, he was on the Faculty in the Department of Mathematics
and Statistics at the University of Massachusetts Amherst. Since 2019 he has been on the Faculty in the Department of Mathematics at the University of Maryland, College Park, where he is currently a Professor. His research interests include graph matching, statistical
inference on random graphs, pattern recognition, dimensionality reduction, stochastic processes, and high-dimensional data analysis.
\end{IEEEbiography}

\newpage

\onecolumn
\section{Appendix}

Herein we collect the proofs of the major results in the manuscript. In the sequel, we write that a sequence of events $(E_n)$ holds with high probability (abbreviated whp) if $\mathbb{P}(E_n)\geq 1-n^{-2}$ for all $n$ sufficiently large.

\subsection{Proof of Theorem \ref{thm:1}}
\label{sec:pf1}

In the proof of Theorem \ref{thm:1}, we will assume the following for our base (non-augmented) RDPG sequence throughout.
\vspace{2mm}

\begin{assumption} \cite{athreya_survey}
\label{ass:1}
Let $\left(\bA_n \sim\text{RDPG}(\bX_n)\right)_{n=2}^\infty$ be a sequence of random dot product graphs with $\bA_n$ being the $n \times n$ adjacency matrix. 
We will assume that
\begin{itemize}
\item[i.] $\bX_n$ is of rank $d$ for all $n$ sufficiently large;
\item[ii.] There exists constant $a > 0$ such that for all $n$ sufficiently large,
\begin{equation*}
    \delta(\bP_n) := \max_{i}\sum_{j=1}^{n}(\bP_n)_{ij} \geq \log^{4+a}(n)
\end{equation*}
\item[iii.] There exists constant $c_0 > 0$ such that for all $n$ sufficiently large,
\begin{equation*}
    \gamma(\bP_n) := \frac{\lambda_d (\bP_n)}{\delta(\bP_n)} \geq c_0
\end{equation*}
where $\lambda_d (\bP_n)$ is the $d$-th largest eigenvalue of $\bP_n$.
\item[iv.] There exists constants $c_1,c_2 > 0$ and a sequence of orthogonal matrices $\widetilde\bW_n$ such that for all $i,j$ and $n$ sufficiently large,
\begin{equation*}
\frac{c_1}{\sqrt{n}}\leq |(\bU_{\bP_n}\widetilde\bW_n)_{i,j}|\leq \frac{c_2}{\sqrt{n}}
\end{equation*}
\end{itemize}
\end{assumption}

\vspace{2mm}

For ease of notation, below we will write $\delta=\delta(\bP_n)$ and $\alpha=\alpha(n)$.
We now consider $\left(\bA_n \sim\text{RDPG}(\bX_n)\right)_{n=2}^\infty$ to be a sequence of random dot product graphs satisfying the conditions of Assumption \ref{ass:1}.
Suppressing the implicit dependence of the parameters on $n$, we have that the edge probability matrix for $\bA$ is given by $\bP = \bX \bX^\top$. 
Similarly, for the augmented RDPG graph $\bAc$, we have $\bAc \sim$ RDPG $(\bXc)$ and the edge probability matrix is given by 
$$\bP^{(c)} = \bXc (\bXc)^\top = \bX \bX^\top + \bVc (\bVc)^\top=\bP+\bVc (\bVc)^\top$$ 
where we augment the $\bX$ matrix with an additional column $\bVc$ to maximally increase the probability of connections between vertices in a set $\mathcal{C}$ (i.e., to maximally increase the probability of an edge forming between every pair of vertices in $\mathcal{C}$).
In the following proof, we will assume that $\alpha=\|\bVc\|_0 =o(\delta^{5/6})$.

Note that $\alpha=\|\bVc\|_0$
count the number of non-zero entries in $\bVc$.
We have that entrywise $0\leq \bVc \leq 1 $ so that for each $i\in[n]$, $ 0 \leq (\bVci)^2\leq\bVci\leq 1$, and
\begin{align*}
 \|\bVc (\bVc)^\top \|  &= \|\bVc \|_2 \|\bVc \|_2 = \|\bVc \|_2^2  = \sum_{i=1}^{n} (\bVci)^2 \leq \alpha 
\end{align*}
Next note that
 $\bAc - \bX\bX^\top = \bAc - \bXc\bXct + \bXc\bXct-  \bX\bX^\top $, so that
\begin{align*}
    \| \bAc - \bX\bX^\top \|  \leq \| \bAc - \bXc\bXct \| + \|\bVc \bVct \|.
\end{align*}
Theorem 21 in \cite{athreya_survey} (refined spectral norm control of $\bA - \bP$, adapted there from the analogous result in \cite{lu13:_spect}) then yields that $\whp$, we have 
$\| \bAc - \bXc\bXct \|$ is of order $O(\sqrt{\delta(\bPc_n)})$. 
Note that
$\delta(\bPc_n)\leq \delta+\alpha=O(\delta)$ so that  
$\| \bAc - \bXc\bXct \|$ is of order $O(\sqrt{\delta})$ $\whp$,
implying that $\| \bAc - \bX\bX^\top \| = O(\sqrt{\delta}+\alpha)$ $\whp$. 

In what follows, we will adapt the proof of \cite[Theorem 26]{athreya2017statistical} to the present setting, accounting for $\alpha$ at each stage as appropriate.

By the Davis-Kahan Theorem (see, for example, \cite{DK_usefulvariant}), letting $\bU_{\bP}$ (resp., $\bU_{\bAc}$, , with ordered eigenvalues in $\bS_{\bAc}$) be the matrix with columns composed of the $d$-largest eigenvectors of $\bP$ (resp., $\bAc$),
we have that there exists a constant $C>0$ (that can change line--to--line) such that (where $\theta_i$ are the principal angles between the subspaces spanned by $\bU_{\bAc}$ and $\bU_{\bP}$)
\begin{align*}
\|  \bU_{\bAc}\bU_{\bAc}^\top -\bU_{\bP} \bU_{\bP}^\top    \| 
= \max_{i} \|\sin{\theta_i}\| 
&\leq\sqrt{d}\frac{C \|\bAc - \bP \|}{\lambda_d(\bP)}\\
&\leq C\sqrt{d}\left(\frac{1}{\delta^{1/2}}+\frac{\alpha}{\delta}\right)
\end{align*}
The variant of the Davis-Kahan theorem given in \cite{rohe2011spectral} yields also that there is a constant $C$ and an orthonormal matrix $\bW \in \RR^{d \times d} $ such that 
\begin{equation}
\label{eq:DK2}
    \|\bU_{\bP}\bW - \bU_{\bAc} \|_F \leq C\sqrt{d}\frac{\|\bAc - \bP \|}{\lambda_d(\bP)}\leq C\sqrt{d}\left(\frac{1}{\delta^{1/2}}+\frac{\alpha}{\delta}\right)
\end{equation}

\noindent 
Letting $\bW_1 \bSigma \bW_2^\top$ be the singular value decomposition of $\bU_{\bP}^\top\bU_{\bAc}$ (with singular values denoted $\sigma_i$), 
we have that w.h.p. 
\begin{align*}
    \| \bU_{\bP}^\top\bU_{\bAc} - \underbrace{\bW_1\bW_2^\top}_{:=\bW^*} \| & = 
    \| \bSigma - I\| = \max_{i}|1-\sigma_i| \leq \max_{i}(1-\sigma_i^2) \\
    &= \max_{i}\sin^2(\theta_i) = \|\bU_{\bAc}\bU_{\bAc}^\top -\bU_{\bP} \bU_{\bP}^\top  \|^2\\
    & = 
    O\left(\frac{\delta+\alpha^2}{\delta^{2}}\right)
\end{align*}

We next establish the analogue of Lemma 49 in \cite{athreya2017statistical}, and adopt the notation used therein.
Let $\bR = \bU_{\bAc} - \bU_{\bP}\bU_{\bP}^\top\bU_{\bAc}$, and note that by Eq \ref{eq:DK2}, (as $\bR$ denotes the residual after projection of $\bU_{\bAc}$ onto the column space of $\bU_\bP$)
$$
\| \bU_{\bAc} - \bU_{\bP}\bU_{\bP}^\top\bU_{\bAc} \|_F \leq \min_{\bW\in\mathcal{O}_d} \| \bU_{\bAc} - \bU_\bP\bW  \|_F=O\left(\frac{\delta^{1/2}+\alpha}{\delta}\right)
$$
Next, we note that
\begin{align*}
    &\bW^*\bS_{\bAc} \\
    & = (\bW^* - \bU_\bP^\top \bU_{\bAc})\bS_{\bAc} + \bU_\bP^\top \bU_{\bAc}\bS_{\bAc} \\
    & = (\bW^* - \bU_\bP^\top \bU_{\bAc})\bS_{\bAc} + \bU_\bP^\top(\bAc - \bP)\bU_{\bAc} + \bU_\bP^\top \bP\bU_{\bAc} \\
    &= (\bW^* - \bU_\bP^\top \bU_{\bAc})\bS_{\bAc} + \bU_\bP^\top(\bAc - \bP)\bR +\bU_\bP^\top(\bAc - \bP)\bU_\bP\bU_\bP^\top\bU_{\bAc} + \bU_\bP^\top \bP\bU_{\bAc} \\
    & = (\bW^* - \bU_\bP^\top \bU_{\bAc})\bS_{\bAc} + \bU_\bP^\top(\bAc - \bP)\bR +\bU_\bP^\top(\bAc - \bP)\bU_\bP\bU_\bP^\top\bU_{\bAc} + \bS_\bP\bU_\bP^\top \bU_{\bAc}
\end{align*}
Write $\bS_\bP\bU_\bP^\top \bU_{\bAc} = \bS_\bP(\bU_\bP^\top \bU_{\bAc} - \bW^*) + \bS_\bP\bW^*$; we then rearrange terms above to get
\begin{align*}
\bW^* \bS_{\bAc} - \bS_\bP \bW^*  &= (\bW^*-\bU_\bP^\top \bU_{\bAc})\bS_{\bAc} + \bU_\bP^\top(\bAc - \bP)\bR+\bU_\bP^\top(\bAc - \bP)\bU_\bP\bU_\bP^\top\bU_{\bAc} \\
&+ \bS_\bP(\bU_\bP^\top \bU_{\bAc} - \bW^*).
\end{align*}
Note that by Weyl's Theorem \cite{horn85:_matrix_analy}, we have that for any $i\in[n]$, 
\begin{align*}
    | \lambda_i(\bAc) - \lambda_i(\bP) | &\leq \| \bAc - \bP   \| =O(\sqrt{\delta}+\alpha)=o(\delta^{5/6}).
\end{align*}
and hence both $\|\bS_\bP\|$ and $\|\bS_{\bAc}\|$ are $O(\delta)$. 
Then we can obtain
\begin{align*}
&\|\bW^* \bS_{\bAc} - \bS_\bP \bW^*\|_F \leq \|\bW^*-\bU_\bP^\top \bU_{\bAc} \|_F(\|\bS_{\bAc}\| + \|\bS_\bP \|) + 
\|(\bAc - \bP)\| \cdot\|\bR \|_F \\
 &\quad\quad\quad\quad\quad\quad\quad+ \|\bU_\bP^\top(\bAc - \bP)\bU_\bP\bU_\bP^\top\bU_{\bAc} \|_F \\
 &\leq 
 O\left(1+\frac{\alpha^2}{\delta}\right)
 +\|\bU_\bP^\top(\bAc - \bP)\bU_\bP \|_F \| \bU_\bP^\top\bU_{\bAc} \| \\
 &\leq  O\left(1+\frac{\alpha^2}{\delta}\right)+ 
 \|\bU_\bP^\top(\bAc - \bPc + \bPc -\bP)\bU_\bP \|_F \| \bU_\bP^\top\bU_{\bAc} \|\\
 &\leq  O\left(1+\frac{\alpha^2}{\delta}\right)+ \bigg[\|\bU_\bP^\top(\bAc - \bPc)\bU_\bP\|_F + 
 \|\bU_\bP^\top(\bPc - \bP)\bU_\bP\|_F
 \bigg] \underbrace{\| \bU_\bP^\top\bU_{\bAc} \|}_{\leq 1}
\end{align*}
Now, the localization assumption gives us that
\begin{align*}
\|\bU_\bP^\top(\bPc - \bP)\bU_\bP\|_F&=\|\widetilde\bW^\top\bU_\bP^\top(\bPc - \bP)\bU_\bP\widetilde\bW\|_F\\
&=\sqrt{\sum_{ij}\left(\sum_{kl}(\widetilde\bW^\top\bU_\bP^\top)_{ik}(\bPc - \bP)_{kl}(\bU_\bP\widetilde\bW)_{lj} \right)^2}\\
&\leq\sqrt{\sum_{ij}\left(\sum_{kl}\frac{c_2}{n}(\bPc - \bP)_{kl} \right)^2}\\
&\leq\sqrt{\sum_{ij}\frac{c_2^2\alpha^4}{n^2} }\\
&=O\left(\frac{\alpha^2}{n} \right)
\end{align*}
As in the argument in the proof of Lemma 49 in \cite{athreya_survey}, 
we know $\bU_\bP^\top(\bAc - \bPc)\bU_\bP$ is a $d \times d$ matrix. 
By utilizing Hoeffding's inequality, we get that each entry of $\bU_\bP^\top(\bAc - \bPc)\bU_\bP$ is of order $O(\log n)$ with high probability. 
As a consequence, 
$$
\|\bU_\bP^\top(\bAc - \bPc)\bU_\bP\|_F = O(\log(n))
$$
with high probability, and hence $\whp$ (recalling the assumption that $\alpha=o(\delta^{5/6})$)
\begin{align}
\label{eq:flipeigs}
\|\bW^* \bS_{\bAc} - \bS_\bP \bW^*\|_F=O\left(\log(n)+\frac{\alpha^2}{\delta}\right)
\end{align} 
To establish that
\begin{align}
\label{eq:flipeigs2}
\|\bW^* \bS_{\bAc}^{1/2} - \bS_{\bP}^{1/2} \bW^* \|_F = O\left(\frac{\log n}{\delta^{1/2} }+\frac{\alpha^2}{\delta^{3/2}}\right),
\end{align}
we note that the $ij-$th entry of $\bW^* \bS_{\bAc}^{1/2} - \bS_{\bP}^{1/2} \bW^*$ can be written as (noting $i,j\in[d]$)
$$
\bW^*_{ij}(\lambda_i^{1/2}(\bAc)-\lambda_j^{1/2}(\bP)) =\bW^*_{ij}\frac{\lambda_i(\bAc)-\lambda_j(\bP)}{\lambda_i^{1/2}(\bAc)+\lambda_j^{1/2}(\bP)}
$$
where $\lambda_1 \geq \lambda_2\geq \dots \geq \lambda_n$ are the ordered eigenvalues of the matrix, and we recall that Weyl's theorem gives the denominator is of order $\delta^{1/2}(\bP)$.

We next establish the analogue of 
Theorem 50 in \cite{athreya_survey}.
Adopting the notation used therein, let $\bR_1 = \bU_\bP\bU_\bP^\top \bU_{\bAc} - \bU_\bP\bW^*$ and $\bR_2 = \bW^* \bS_{\bAc}^{1/2} - \bS_{\bP}^{1/2} \bW^*$. 
We deduce that
\begin{align*}
    \hbXc &- \bU_\bP\bS_\bP^{1/2}\bW^* = \bU_{\bAc}\bS_{\bAc}^{1/2}-\bU_\bP\bW^*\bS_{\bAc}^{1/2} + \bU_\bP(\bW^*\bS_{\bAc}^{1/2}- \bS_{\bP}^{1/2} \bW^*) \\
    &= \bU_{\bAc}\bS_{\bAc}^{1/2} - \bU_\bP\bU_\bP^\top \bU_{\bAc}\bS_{\bAc}^{1/2} + \bR_1\bS_{\bAc}^{1/2}+\bU_\bP\bR_2 \\
    &= (\bAc-\bP)\bU_{\bAc}\bS_{\bAc}^{-1/2} - \bU_\bP\bU_\bP^\top (\bAc-\bP)\bU_{\bAc}\bS_{\bAc}^{-1/2}+\bR_1\bS_{\bAc}^{1/2}+\bU_\bP\bR_2 
\end{align*}
since $\bU_\bP\bU_\bP^\top \bP=\bP$ and $\bU_{\bAc}\bS_{\bAc}^{1/2}=\bAc\bU_{\bAc}\bS_{\bAc}^{-1/2}$.
Let 
\begin{align}
\label{eq:R3}
    \bR_3 = \bU_{\bAc}-\bU_\bP\bW^* = \bU_{\bAc} - \bU_\bP\bU_\bP^\top\bU_{\bAc} + \bR_1,
\end{align} 
we can write
\begin{align}
    \hbXc - \bU_\bP\bS_\bP^{1/2}\bW^* 
    =& (\bAc-\bP)\bU_\bP\bW^*\bS_{\bAc}^{-1/2}\label{eq:5terms1}\\  
    &- \bU_\bP\bU_\bP^\top (\bAc-\bP)\bU_\bP\bW^*\bS_{\bAc}^{-1/2}\label{eq:5terms2}\\
    &+ (\bI-\bU_\bP\bU_\bP^\top)(\bAc-\bP)\bR_3\bS_{\bAc}^{-1/2}\label{eq:5terms3}\\ 
    &+ \bR_1\bS_{\bAc}^{1/2}\label{eq:5terms4}\\ 
    &+\bU_\bP\bR_2\label{eq:5terms5}
\end{align}
We will bound the above pieces in Eq. \ref{eq:5terms1}-\ref{eq:5terms5} one at a time.

\subsubsection{Bounding Eq. \ref{eq:5terms4}-\ref{eq:5terms5}}
% *****

Now, combining the above bounds we have that $\whp$ (again using the assumption that $\alpha=o(\delta^{3/4}(\bP))$ and the delocalization assumption)
\begin{align*}
    \|\bR_1 \|_{2\rightarrow\infty} & = \|\bU_\bP\widetilde\bW (\widetilde\bW^\top\bU_\bP^\top \bU_{\bAc} - \widetilde\bW^\top\bW^*)\|_{2\rightarrow\infty}\\
    &\leq \|\bU_\bP\widetilde\bW\|_{2\rightarrow\infty}\|\widetilde\bW^\top\bU_\bP^\top \bU_{\bAc} - \widetilde\bW^\top\bW^*\|\\
    &=O\left(\frac{\delta+\alpha^2}{\sqrt{n}\delta^{2}}\right)
    \end{align*}
This provides that the term in \ref{eq:5terms4} is of order
$$
\|\bR_1\bS_{\bAc}^{1/2}\|_{2\rightarrow\infty}=O\left(\frac{\delta+\alpha^2}{\sqrt{n}\delta^{3/2}}\right).
$$
Delocalization also provides that the term in \ref{eq:5terms5} is of order
\begin{align*}
    \|\bU_\bP\bR_2 \|_{2\rightarrow\infty} &\leq \|\bU_\bP\widetilde\bW\|_{2\rightarrow\infty}\|\widetilde\bW^\top\bR_2\|\\
    &=O\left(\frac{\log n}{\sqrt{n}\delta^{1/2}}+\frac{\alpha^2 }{\sqrt{n}\delta^{3/2}}\right).
\end{align*}

\subsubsection{Bounding Eq. \ref{eq:5terms2}}
%******

Again applying Hoeffding's inequality, the delocalization assumption, and Weyl's Theorem, we have that 
\begin{align*}
    \|\bU_\bP\bU_\bP^\top &(\bAc-\bP)\bU_\bP\bW^*\bS_{\bAc}^{-1/2}\|_{2\rightarrow\infty}\\
    &\leq  \|\bU_\bP\widetilde\bW\|_{2\rightarrow\infty}\|\widetilde\bW^\top\bU_\bP^\top (\bAc-\bPc +\bPc - \bP)\bU_\bP\bW^*\bS_{\bAc}^{-1/2} \| \\
    & \leq \|\bU_\bP\widetilde\bW\|_{2\rightarrow\infty}(\|\bU_\bP^\top (\bAc-\bPc)\bU_\bP\|_F  + \| \bU_\bP^\top (\bPc - \bP)\bU_\bP\|_F)\|\bS_{\bAc}^{-1/2}\|_F \\
    & = O\left(\frac{\log n}{ \sqrt{n}\delta^{1/2}}+\frac{\alpha^2}{n^{3/2}\delta^{1/2}}\right)
\end{align*}

\subsubsection{Bounding Eq. \ref{eq:5terms1}}

We next turn our attention to bounding 
$(\bAc-\bP)\bU_\bP\bW^*\bS_{\bAc}^{-1/2}$.  
Note that
\begin{align*}
\|(\bAc-\bP)&\bU_\bP\bW^*\bS_{\bAc}^{-1/2}\|_{2\rightarrow\infty} \\
&\leq\left\| (\bAc-\bP)\bU_\bP\bS_\bP^{-1/2}\bW^*\right\|_{2\rightarrow\infty} + \left\|(\bAc-\bP)\bU_\bP (\bS_\bP^{-1/2}\bW^* -\bW^*\bS_{\bAc}^{-1/2} ) \right\|_{2\rightarrow\infty}
\end{align*}
As in the argument used to prove Eq. \ref{eq:flipeigs2}, we derive that $\whp$
$$
\|\bW^* \bS_{\bAc}^{-1/2} - \bS_{\bP}^{-1/2} \bW^* \| = O\left(\frac{\log n}{\delta^{3/2}}+\frac{\alpha^2 }{\delta^{5/2}}\right)
$$
Therefore
\begin{align}
\label{eq:eq51}
\left\|(\bAc-\bP)\bU_\bP (\bS_\bP^{-1/2}\bW^* -\bW^*\bS_{\bAc}^{-1/2} ) \right\|_{2\rightarrow\infty}=O\left(\frac{\log n}{\delta}+\frac{\alpha^2 }{\delta^2}+\frac{\alpha\log n}{\delta^{3/2}}+\frac{\alpha^3 }{\delta^{5/2}}  \right)
\end{align}
For the remaining term, we have that
\begin{align}
\left\| (\bAc-\bP)\bU_\bP\bS_\bP^{-1/2}\bW^*\right\|_{2\rightarrow\infty}&=\left\| (\bAc-\bP)\bU_\bP\widetilde \bW\right\|_{2\rightarrow\infty}\|\bS_\bP^{-1/2}\|\notag \\&\leq \|\bS_\bP^{-1/2}\|\left( \left\| (\bPc-\bP)\bU_\bP\widetilde \bW\right\|_{2\rightarrow\infty}+\left\| (\bAc-\bPc)\bU_\bP\widetilde \bW\right\|_{2\rightarrow\infty} \right)\label{eq:eq52}
\end{align}
For a given $i$, we can use Hoeffding's inequality to show that the $j-$th element of the vector 
\begin{align}
\label{eq:Hoeff2}
\left(\left((\bAc-\bPc)\bU_\bP\widetilde\bW\right)_{i,\bullet}\right)_j=\sum_k (\bAc_{i,k}- \bPc_{i,k})(\bU_\bP\widetilde\bW)_{kj}
\end{align}
is $O(\log n)$ $\whp$
Next note that if vertex $i$ is not an pseudo-clique vertex, then $(\bPc - \bP)_{i,\bullet}=\vec 0$, and so 
$$
\left((\bPc - \bP)\bU_\bP\widetilde\bW\right)_{i,\bullet}=(\bPc - \bP)_{i,\bullet}(\bU_\bP\widetilde\bW)=\vec 0.
$$
If $i$ is a pseudo-clique vertex,
the delocalization assumption yields that 
\begin{align*}
\left|\left(\left((\bPc - \bP)\bU_\bP\widetilde\bW\right)_{i,\bullet}\right)_j\right| &=\left|\sum_k (\bPc_{ik} - \bP_{ik})(\bU_\bP\widetilde\bW)_{kj}\right|\\
&\leq \sum_k \left|(\bPc_{ik} - \bP_{ik})(\bU_\bP\widetilde\bW)_{kj}\right|\\
&\leq \frac{\alpha}{\sqrt{n}}
\end{align*} 
Combining the above with a union bound over $i$ and $j$ (for Eq. \ref{eq:Hoeff2}), we have that $\whp$
$$\left\| (\bAc-\bP)\bU_\bP\bS_\bP^{-1/2}\bW^*\right\|_{2\rightarrow\infty}
=O\left(\frac{\alpha^2 }{\delta^2}+
\frac{\alpha^3 }{\delta^{5/2}} +\frac{\log n}{\delta^{1/2}}+\frac{\alpha}{\sqrt{\delta n}}  \right)
$$

\subsubsection{Bounding Eq. \ref{eq:5terms3}}
%******

Next note that the decomposition in Theorem 3.1 of \cite{cape2toinfty} allows us to decompose $\bR_3= \bU_{\bAc}-\bU_\bP\bW^*$ via
\begin{align}
\bU_{\bAc}-\bU_\bP\bW^*=&(\mathbf{I}-\bU_\bP\bU_\bP^T)(\bA^{(c)}-\bP)\bU_\bP\bW^*\bS_{\bA^{(c)}}^{-1}\label{eq:r31}\\ 
&+(\mathbf{I}-\bU_\bP\bU_\bP^T)(\bA^{(c)}-\bP)(\bU_{\bA^{(c)}}-\bU_\bP\bW^*)\bS_{\bA^{(c)}}^{-1}\label{eq:r32}\\ 
&+\bU_\bP(\bU_\bP\bU_{\bAc}-\bW^*)\label{eq:r33}
\end{align}
The term involving $\bR_3$ then becomes 
\begin{align}
&(\mathbf{I}-\bU_\bP\bU_\bP^T)(\bA^{(c)}-\bP)(\mathbf{I}-\bU_\bP\bU_\bP^T)(\bA^{(c)}-\bP)\bU_\bP\bW^*\bS_{\bA^{(c)}}^{-3/2}\label{eq:r311}\\ 
&+(\mathbf{I}-\bU_\bP\bU_\bP^T)(\bA^{(c)}-\bP)(\mathbf{I}-\bU_\bP\bU_\bP^T)(\bA^{(c)}-\bP)(\bU_{\bA^{(c)}}-\bU_\bP\bW^*)\bS_{\bA^{(c)}}^{-3/2}\label{eq:r321}\\ 
&+(\mathbf{I}-\bU_\bP\bU_\bP^T)(\bA^{(c)}-\bP)\bU_\bP(\bU_\bP\bU_{\bAc}-\bW^*)\bS_{\bA^{(c)}}^{-1/2}\label{eq:r331}
\end{align}
We will bound each of these in turn.
For \ref{eq:r331}, we have that with high probability
\begin{align*}
\|(\mathbf{I}-\bU_\bP&\bU_\bP^T)(\bA^{(c)}-\bP)\bU_\bP(\bU_\bP\bU_{\bAc}-\bW^*)\bS_{\bA^{(c)}}^{-1/2}\|_{2\rightarrow\infty}
\\ 
\leq& \underbrace{\|(\bA^{(c)}-\bP)\bU_\bP\widetilde{\bW}\|_{2\rightarrow\infty}\|(\bU_\bP\bU_{\bAc}-\bW^*)\bS_{\bA^{(c)}}^{-1/2}\|}_{=O\left(\frac{(\sqrt{n}\log n+\alpha )(\delta+\alpha^2)}{\sqrt{n}\delta^{5/2}}\right)}\\ 
&+\underbrace{\|\bU_\bP\widetilde\bW\|_{2\rightarrow\infty}\|\bU_\bP^T(\bA^{(c)}-\bP)\bU_\bP\|\cdot\|(\bU_\bP\bU_{\bAc}-\bW^*)\bS_{\bA^{(c)}}^{-1/2}\|}_{
=O\left(\frac{(n\log n+\alpha^2)(\delta+\alpha^2)}{n^{3/2}\delta^{5/2}}
\right)}\\ 
=&O\left(
\frac{\log n}{\delta^{3/2}}+
\frac{\alpha}{\sqrt{n}\delta^{3/2}}+
\frac{\alpha^2\log n}{\delta^{5/2}} 
+
\frac{\alpha^3}{\sqrt{n}\delta^{5/2}}\right).
\end{align*}
For \ref{eq:r321}, we have that with high probability
\begin{align*}
\|(\mathbf{I}-&\bU_\bP\bU_\bP^T)(\bA^{(c)}-\bP)(\mathbf{I}-\bU_\bP\bU_\bP^T)(\bA^{(c)}-\bP)(\bU_{\bA^{(c)}}-\bU_\bP\bW^*)\bS_{\bA^{(c)}}^{-3/2}\|_{2\rightarrow\infty}\\ 
&=O\left( \frac{\delta^{3/2}+\alpha^3  }{ \delta^{5/2} } \right)
\end{align*}
For \ref{eq:r311}, we consider
\begin{align}
(\mathbf{I}-\bU_\bP&\bU_\bP^T)(\bA^{(c)}-\bP)(\mathbf{I}-\bU_\bP\bU_\bP^T)(\bA^{(c)}-\bP)\bU_\bP\bW^*\bS_{\bA^{(c)}}^{-3/2}\notag\\ 
=&(\bA^{(c)}-\bP)^2\bU_\bP\bW^*\bS_{\bA^{(c)}}^{-3/2}\label{eq:r312}\\ 
&- \bU_\bP\bU_\bP^T(\bA^{(c)}-\bP)(\mathbf{I}-\bU_\bP\bU_\bP^T)(\bA^{(c)}-\bP)\bU_\bP\bW^*\bS_{\bA^{(c)}}^{-3/2}\label{eq:r322}\\ 
&- (\bA^{(c)}-\bP)\bU_\bP\bU_\bP^T(\bA^{(c)}-\bP)\bU_\bP\bW^*\bS_{\bA^{(c)}}^{-3/2}\label{eq:r332}
\end{align}
The $2\rightarrow\infty$ norm of \ref{eq:r332} can be bounded via 
$$O\left(
\frac{\log^2n}{\delta^{3/2}}+
\frac{\log n\alpha}{\sqrt{n}\delta^{3/2}}+\frac{\alpha^3}{n^{3/2}\delta^{3/2}}
+\frac{\log n\alpha^2}{n\delta^{3/2}}\right),$$ and that of \ref{eq:r322} is bounded via $O\left(\frac{\delta+\alpha^2}{\sqrt{n}\delta^{3/2}}\right)$.
For term \ref{eq:r312}, writing $(\bAc-\bP)=(\bAc-\bPc+\bPc-\bP)$, we have that
\begin{align}
(\bA^{(c)}-\bP)^2\bU_\bP\bW^*\bS_{\bA^{(c)}}^{-3/2}=&(\bA^{(c)}-\bPc)^2\bU_\bP\bW^*\bS_{\bA^{(c)}}^{-3/2}\label{eq:R313}\\ 
&+(\bA^{(c)}-\bPc)(\bPc-\bP)\bU_\bP\bW^*\bS_{\bA^{(c)}}^{-3/2}\label{eq:R323}\\ 
&+(\bPc-\bP)(\bA^{(c)}-\bPc)\bU_\bP\bW^*\bS_{\bA^{(c)}}^{-3/2}\label{eq:R333}\\ 
&+(\bP-\bPc)^2\bU_\bP\bW^*\bS_{\bA^{(c)}}^{-3/2}\label{eq:R343}
\end{align}
Term \ref{eq:R313} has spectral norm of order $O(\delta^{-1/2})$.
For terms \ref{eq:R323}-\ref{eq:R333}, we note (term \ref{eq:R333} being analogous)
\begin{align*}
\|(\bA^{(c)}-\bPc)(\bPc-\bP)\bU_\bP\bW^*\bS_{\bA^{(c)}}^{-3/2}\|_{2\rightarrow\infty}\leq 
\|(\bA^{(c)}-\bPc)(\bPc-\bP)\bU_\bP\widetilde \bW\|_{2\rightarrow\infty}\|\bS_{\bA^{(c)}}^{-3/2}\|_{2\rightarrow\infty}
\end{align*}
Now 
$$\left[(\bA^{(c)}-\bPc)(\bPc-\bP)\right]_{ij}=\begin{cases}
0&\text{ if }j\text{ is not a pseudo-clique vertex }\\ 
O(\alpha)&\text{ if }j\text{ is a pseudo-clique vertex }
\end{cases},$$
and each row of $(\bA^{(c)}-\bPc)(\bPc-\bP)$ has $O(\alpha)$ nonzero entries each of $O(\alpha)$.
Therefore 
$$\left[(\bA^{(c)}-\bPc)(\bPc-\bP)\bU_\bP\widetilde \bW\right]_{ij}=O(\alpha^2/\sqrt{n}),$$
 and
 \begin{align*}
\|(\bA^{(c)}-\bPc)(\bPc-\bP)\bU_\bP\widetilde \bW\|_{2\rightarrow\infty}\|\bS_{\bA^{(c)}}^{-3/2}\|_{2\rightarrow\infty}=O\left( \frac{\alpha^2}{\sqrt{n}\delta^{3/2}} \right)
\end{align*}.
Turning our attention to term \ref{eq:R343}, we see that
\begin{align*}
\left\|(\bP-\bPc)^2\bU_\bP\bW^*\bS_{\bA^{(c)}}^{-3/2}\right\|_{2\rightarrow\infty}&\leq \left\|(\bP-\bPc)^2\bU_\bP\widetilde\bW\right\|_{2\rightarrow\infty} \|\bS_{\bA^{(c)}}^{-3/2}\|=O\left( \frac{\alpha^2}{\sqrt{n}\delta^{3/2}} \right)
\end{align*}

Combining the above, we have that

$$
(\bI-\bU_\bP\bU_\bP^\top)(\bAc-\bP)\bR_3\bS_{\bAc}^{-1/2}=O\left(
\frac{\alpha^2\log n}{\delta^{5/2}} 
+
\frac{
\alpha^3  }{ \delta^{5/2} }
+\delta^{-1/2}
 +\frac{\alpha^2}{\sqrt{n}\delta^{3/2}}
\right).
$$

\subsubsection{Proof of desired bound}

Combining the above, we have that \whp, 
$$
    \| \hbXc - \bU_\bP\bS_\bP^{1/2}\bW^*\|_{2\rightarrow\infty}= 
O\left(
\frac{\alpha^3  }{ \delta^{5/2} }+
\frac{\alpha^2 }{\delta^2}+
\frac{\log n}{\delta^{1/2}}+
\frac{\alpha}{\sqrt{\delta n}}
\right)
$$ 
as desired.
The result then follows by combining the above inequality with the analogous (significantly tighter) concentration bound for the ASE of $\bA$ into $\mathbb{R}^d$ as found in \cite[Theorem 26]{athreya2017statistical}.

\subsection{Proof of Theorem \ref{thm:lwb}}
\label{sec:pflwb}
Consider the setting where $\delta=\Theta(n)$, $\alpha=\omega(n^{1/2}\log n)$ and $\alpha=o(n^{3/4})$,  
so that the lead order term in 
the bound of $
    \| \hbXc - \bU_\bP\bS_\bP^{1/2}\bW^*\|_{2\rightarrow\infty}$ 
is $O(\frac{\alpha}{n})$
Note that in the expansion of $\hbXc - \bU_\bP\bS_\bP^{1/2}\bW^*$ is Eqs. \ref{eq:5terms1}--\ref{eq:5terms5}, we have that 
Eq. \ref{eq:5terms4} is of order
$$
% 8
O\left(\frac{\delta+\alpha^2}{\sqrt{n}\delta^{3/2}}\right)
$$
which is $o(\alpha/n)$; 
Eq. \ref{eq:5terms5} is of order
$$
%9
O\left(\frac{\log n}{\sqrt{n}\delta^{1/2}}+
\frac{\alpha^2 }{\sqrt{n}\delta^{3/2}}\right)
$$
which is $o(\alpha/n)$;
Eq. \ref{eq:5terms3} is of order
$$
%7
O\left(\frac{\alpha^2\log n}{\delta^{5/2}} +
\frac{\alpha^3  }{ \delta^{5/2} }+
 \delta^{-1/2}+
 \frac{\alpha^2}{\sqrt{n}\delta^{3/2}}\right)
$$
which is $o(\alpha/n)$;
Eq. \ref{eq:5terms2} is of order
$$
%6
O\left(\frac{\log n}{ \sqrt{n}\delta^{1/2}}+
 \frac{\alpha^2}{n^{3/2}\delta^{1/2}}\right)
 $$
which is $o(\alpha/n)$.
Lastly, in Eq. \ref{eq:5terms2}, the term in Eq. \ref{eq:eq51} is of order
$$
O\left(\frac{\log n}{\delta}+\frac{\alpha^2 }{\delta^2}+\frac{\alpha\log n}{\delta^{3/2}}+\frac{\alpha^3 }{\delta^{5/2}}  \right)
$$
which is $o(\alpha/n)$;
The second term in Eq. \ref{eq:eq52} is of order $O(\log n/\delta^{1/2})$ which is $o(\alpha/n)$.
The first term in Eq. \ref{eq:eq52} is then the key term.
If $i$ is not a pseudo-clique vertex, then 
\begin{align*}
\left[(\bP-\bPc)\bU_\bP\bS_{\bP}^{-1/2}\bW^*\right]_{i\bullet}=\left[(\bP-\bPc)\right]_{i\bullet}\bU_\bP\bS_{\bP}^{-1/2}\bW^*=\vec{0}.
\end{align*}
If $i$ is a pseudo-clique vertex that is in alignment with $\bP$, we have that 
\begin{align*}
\left\| \left[(\bP-\bPc)\bU_\bP\bS_{\bP}^{-1/2}\bW^*\right]_{i\bullet}\right\|\geq \left\| \left[(\bP-\bPc)\bU_\bP\check\bW\right]_{i\bullet}\right\| 
\underbrace{\sigma_{\min}(\bS_{\bP}^{-1/2})}_{=\Theta(\delta^{-1/2})},
\end{align*}
and that
\begin{align*}
\left[(\bP-\bPc)\bU_\bP\check \bW\right]_{ij}&=
\sum_{k}(\bPc-\bP)_{ik}(\bU_\bP\check \bW)_{k,j}.
\end{align*}
The assumption that the pseudo-clique indexed entries of $\bV$ are $\Theta(1)$ combined with the congruence assumption provides that the pseudo-clique indexed entries of $(\bU_\bP\check \bW)_{k,j}$ are all of the same sign and order ($\Theta(n^{-1/2})$), and so 
$$
\left[(\bP-\bPc)\bU_\bP\check \bW\right]_{ij}^2=\Theta(\alpha^2/n)
$$
and hence
\begin{align}
\label{eq:lwrbnd}
\left\| \left[(\bP-\bPc)\bU_\bP\bS_{\bP}^{-1/2}\bW^*\right]_{i\bullet}\right\|=\Omega\left( \frac{\alpha}{\sqrt{n\delta} }\right)
\end{align}
as desired.

\subsection{Proof of Theorem \ref{thm:2}}
\label{sec:pf2}

In the proof of Theorem \ref{thm:2}, we will assume the following for our base (non-augmented) RDPG sequence throughout.
\vspace{2mm}

\begin{assumption}
\label{ass:2}
Let $\left(\bA_n \sim\text{RDPG}(\bX_n)\right)_{n=2}^\infty$ be a sequence of random dot product graphs with $\bA_n$ being the $n \times n$ adjacency matrix, and assume the corresponding vertex class vector being provided by $\bY$. 
We will assume that
\begin{itemize}
\item[i.] For each $k\in[K]$, $n_k=\sum_{i=1}^n\mathds{1}\{\bY_i=k\}=\Theta(n)$;
\item[ii.] We require $\xi(n):=\min_{i,k}\sum_{j=1:\bY_j=k}^{n}(\bP_n)_{ij}=\omega(\sqrt{n\log(n)})
$
\end{itemize}
\end{assumption}
\noindent From the definition of the encoder embedding,
\begin{align*}
\bZc_{ik}&=\frac{\sum_{j=1:j\neq i,Y_j=k}^n\bAc_{ij}}{n_k}\\
&=\frac{\sum_{\substack{j=1:j\neq i,Y_j=k,\\\{i,j\}\not\subset\mathcal{C}}}^n\bAc_{ij}}{n_k}+
\underbrace{\frac{\sum_{\substack{j=1:j\neq i,Y_j=k,\\\{i,j\}\subset\mathcal{C}}}^n\bAc_{ij}}{n_k}}_{\leq \alpha/n_k}
\end{align*}
Note that if $\{i,j\}\not\subset\mathcal{C}$, then $\bAc_{ij}$ and $\bA_{ij}$ are identically distributed, and that
\begin{align*}
\mathbb{E}(\bZc_{ik})&=\frac{\sum_{\substack{j=1:j\neq i,Y_j=k,\\\{i,j\}\not\subset\mathcal{C}}}^n\bX_i^\top\bX_j}{n_k}+
\frac{\sum_{\substack{j=1:j\neq i,Y_j=k,\\\{i,j\}\subset\mathcal{C}}}^n\bX_i^\top\bX_j+\bVc_i\bVc_j}{n_k}\\
&=\mathbb{E}(\bZ_{ik})+ \frac{\sum_{\substack{j=1:j\neq i,Y_j=k,\\\{i,j\}\subset\mathcal{C}}}^n\bVc_i\bVc_j}{n_k}
\end{align*}
Note that term $$\left|\frac{\sum_{\substack{j=1:j\neq i,Y_j=k,\\\{i,j\}\subset\mathcal{C}}}^n\bVc_i\bVc_j}{n_k}\right|\leq \alpha/n_k$$ 
for all $i$.
Note that if $\alpha=o(\xi(n))$, then
\begin{align*}
|\{j\text{ s.t. }j\neq i,Y_j=k,\{i,j\}\not\subset\mathcal{C}\}|&=n_k-\mathds{1}\{Y_i=k\}-|\{j\text{ s.t. }j\neq i,Y_j=k,\{i,j\}\subset\mathcal{C}\}|\\
&=n_k(1-o(1)).
\end{align*}
From Hoeffding's inequality, we then have that $T_{ik}:=\sum_{\substack{j=1:j\neq i,Y_j=k,\\\{i,j\}\not\subset\mathcal{C}}}^n\bAc_{ij}$ satisfies
\begin{align*}
\mathbb{P}(|T_{ik}-\mathbb{E}T_{ik}|\geq \sqrt{2n_k\log n})&\leq 2\text{exp}\left\{-\frac{4n_k\log n}{n_k}   \right\}=\frac{2}{n^4}
\end{align*}
We then have that with probability at least
$1-2/n^4$,
\begin{align}
\label{eq:gee1}
\left(\bZc_{ik}-\mathbb{E}(\bZ_{ik})\right)^2&=
\left(\frac{T_{ik}}{n_k}+\frac{\sum_{\substack{j=1:j\neq i,Y_j=k,\\\{i,j\}\subset\mathcal{C}}}^n\bAc_{ij}}{n_k}-\frac{\mathbb{E}T_{ik}}{n_k}-\frac{\sum_{\substack{j=1:j\neq i,Y_j=k,\\\{i,j\}\subset\mathcal{C}}}^n\bX_i^\top\bX_j}{n_k}\right)^2\\
&\leq \frac{4n_k\log n+8\alpha^2}{n_k^2}\notag
\end{align}
An analogous result for $(\bZ_{ik}-\mathbb{E}(\bZ_{ik}))^2$ yields that with probability at least
$1-2/n^4$
\begin{align*}
\left(\bZ_i-\mathbb{E}(\bZ_i)\right)^2
&\leq \frac{4n_k\log n}{n_k^2}
\end{align*}
Summing over $k\in[K]$, and taking the intersection over $i\in[n]$ (a union bound over the complements) yields then that w.h.p. 
\begin{align*}
\max_i \|\bZc_i-\bZ_i\|_2
&\leq \underbrace{\max_i \|\bZc_i-\mathbb{E}(\bZ_i)\|_2}_{=O\left( \frac{(\sqrt{K(n\log n)}+\sqrt{K}\alpha}{n}\right)} +\underbrace{\max_i\|\bZ_i-\mathbb{E}(\bZ_i)\|_2}_{=O\left( \frac{\sqrt{Kn\log n}}{n}\right)}
\end{align*}
as desired.

\subsection{Proof of Theorem \ref{thm:lwb2}}
\label{sec:pflwb2}
In the above theorem, note that if $i$ is not a pseudo-clique vertexx, then $\bZ_{ik}^{(c)}\stackrel{dist.}{=}\bZ_{ik}$.  The upper bound on $\max_{i\notin\mathcal{C}}\|\bZ_i^{(c)}-\bZ_i\|$ follows then verbatim along the lines of the proof of Theorem \ref{thm:2}.
If $i$ is a pseudo-clique vertex, then
letting 
$$\xi_{ik}:=\sum_{\substack{j=1:j\neq i,\\Y_j=k,j\in\mathcal{C}}}\bAc_{ij},$$
we have that $\mathbb{E}\xi_{ik}=\Theta(\alpha)$.
Hoeffding's inequality applied to $\xi_{ik}$ yields that
\begin{align*}
\mathbb{P}(|\xi_{ik}-\mathbb{E}\xi_{ik}|\geq \sqrt{2(\alpha-1)\log n})&\leq 2\text{exp}\left\{-\frac{4(\alpha-1)\log n}{(\alpha-1)}   \right\}=\frac{2}{n^4}.
\end{align*}
From this, we have that probability at least $1-4/n^4$,
\begin{align*}
    \sum_{\substack{j=1:j\neq i,\\Y_j=k,j\in\mathcal{C}}}\!\!\!\!\!\bAc_{ij}-\!\!\!\!\!\sum_{\substack{j=1:j\neq i,\\Y_j=k,j\in\mathcal{C}}}\!\!\!\!\!\bX_i^\top\bX_j=
    \underbrace{\xi_{ik}-\mathbb{E}\xi_{ik}}_{=O(\sqrt{\alpha\log n})}+\underbrace{\sum_{\substack{j=1:j\neq i,\\Y_j=k,j\in\mathcal{C}}}\!\!\!\!\!\bV_i^{(c)}\bV_j^{(c)}}_{=\Theta(\alpha)},
\end{align*}
and an expansion as in Eq. \ref{eq:gee1} yields
\begin{align*}
n_k^2\left(\bZc_{ik}-\mathbb{E}(\bZ_{ik})\right)^2=&
\underbrace{(T_{ik}-\mathbb{E}T_{ik})^2}_{=O(n\log n)}+
\underbrace{2(T_{ik}-\mathbb{E}T_{ik})\left(\sum_{\substack{j=1:j\neq i,\\Y_j=k,j\in\mathcal{C}}}\!\!\!\!\!\bAc_{ij}-\!\!\!\!\!\sum_{\substack{j=1:j\neq i,\\Y_j=k,j\in\mathcal{C}}}\!\!\!\!\!\bX_i^\top\bX_j\right)}_{O(\sqrt{n\alpha\log n}+\alpha\sqrt{n\log n})} 
+\underbrace{\left(\sum_{\substack{j=1:j\neq i,\\Y_j=k,j\in\mathcal{C}}}\!\!\!\!\!\bAc_{ij}-\!\!\!\!\!\sum_{\substack{j=1:j\neq i,\\Y_j=k,j\in\mathcal{C}}}\!\!\!\!\!\bX_i^\top\bX_j\right)^2}_{O(\alpha\log n +\alpha^{3/2}\sqrt{\log n})+\Theta(\alpha^2) }
\end{align*}
given the growth rate on $\alpha$, we have that $\left(\bZc_{ik}-\mathbb{E}(\bZ_{ik})\right)^2=\Theta(\alpha^2)/n^2$, and the result follows.

\newpage

\section{Comparison with existing methods}
\label{app:SOTA}
To identify densely connected substructures within graphs, we evaluated and compared four additional algorithms.
For each method we performed independent $nMC = 50$ simulations and reported the precision, recall and F1 score of true clique vertices and predicted clique vertices in the table \ref{table:algos}, respectively. 
The four methods are as follows.
Goldberg’s exact method \cite{goldberg1984finding} finds the subgraph with maximum average degree by solving a sequence of max-flow subproblems.
Charikar's greedy algorithm \cite{charikar2000greedy} is a fast (linear time) $1/2$-approximation algorithm that iteratively removes the lowest-degree vertex.
Greedy++ \cite{boob2020flowless} refines Charikar’s approach
via a more nuanced degree-based pruning combined with local search refinements; empirically, this has resulted in  excellent, scalable performance.
Lastly, the FISTA-based algorithm \cite{harb2022faster} used a convex relaxation of the densest subgraph problem (solved via the Fast Iterative Shrinkage-Thresholding Algorithm \cite{beck2009fast}); this method works particularly well in the weighted graph setting.
Note that traditional clique detection routines are not appropriate for detecting the pseudo-cliques in this RDPG example, as the pseudo-cliques are likely to have missing edges and are not true cliques.

The evaluation considers the two settings as mentioned in Section \ref{sec:exp}, the true planted clique on $\bA^{(c)}$ and a pseudo-clique structure on $\bX^{(c)}$. Across all methods, the F1 scores increase with larger planted clique sizes. For small cliques (e.g., $n_c = \log(n)$), all methods exhibit low detection accuracy, whereas for denser settings (e.g., $n_c = 0.2n$), the detection accuracy increases significantly.

\begin{table}[h!]
\centering
\begin{tabular}{|c|c|c|c|c|c|c|c|c|c|}
\hline
\multirow{2}{*}{$n_c$}  & \multicolumn{2}{c|}{Goldberg’s } & \multicolumn{2}{c|}{Charikar’s} & \multicolumn{2}{c|}{Greedy++}& \multicolumn{2}{c|}{FISTA} & Metric \\ \cline{2-9}
    & $\bA^{(c)}$ & $\bX^{(c)}$ & $\bA^{(c)}$ & $\bX^{(c)}$ & $\bA^{(c)}$ & $\bX^{(c)}$ & $\bA^{(c)}$ & $\bX^{(c)}$ &\\ \hline
$\log(n)$ & 0.591\%& 0.589\% & 0.591\% &0.589\% &0.591\%& 0.589\% & 0.593\% &0.585\%  & Prec.\\
$\sqrt{n}$ & 3.276\% & 3.240\%& 3.276\% & 3.240\% &3.276\%&3.240\%&3.268\%& 3.233\%  & Prec.\\
$n^{2/3}$ &11.334\% & 11.139\% & 11.334\% & 11.140\% & 11.334\%&11.139\%& 11.344\%& 11.126\%  & Prec.\\
$n^{3/4}$ &20.600\% & 20.381\% &20.599\%&20.381\%&20.600\%&20.381\%&20.592\%&20.371\%  & Prec.\\
$0.2n$ &23.409\% & 23.170\%&23.409\%&23.169\%&23.407\%&23.168\%&23.385\%&23.166\% & Prec.\\
\hline
$\log(n)$ & 85\%& 85\% & 85\% &85\% &85\%& 85\% & 85.33\% &84.33\% & Recall\\
$\sqrt{n}$ & 91.354\% & 90.387\%& 91.354\% & 90.387\% &91.354\%&90.387\%&91.161\%& 90.193\% & Recall\\
$n^{2/3}$ &99.414\% & 97.878\% & 99.414\% & 97.899\% & 99.414\%&97.878\%& 99.596\%& 97.838\% & Recall\\
$n^{3/4}$ &100\% & 100\% &100\%&100\%&100\%&100\%&100\%&100\% & Recall\\
$0.2n$ &100\% & 100\%&100\%&100\%&100\%&100\%&100\%&100\%& Recall\\
\hline
$\log(n)$ & 1.174\%& 1.171\% & 1.174\% &1.171\% &1.174\%& 1.171\% & 1.178\% &1.162\% & F1\\
$\sqrt{n}$ & 6.325\% & 6.256\%& 6.325\% & 6.256\% &6.325\%&6.256\%&6.311\%& 6.242\% & F1\\
$n^{2/3}$ &20.349\% & 20.002\% & 20.349\% & 20.003\% & 20.349\%&20.002\%& 20.368\%& 19.981\% & F1\\
$n^{3/4}$ &34.162\% & 33.86\% &34.161\%&33.86\%&34.162\%&33.86\%&34.151\%&33.846\% & F1\\
$0.2n$ &37.936\% & 37.622\%&37.935\%&37.621\%&37.934\%&37.620\%&37.904\%&37.617\%& F1\\
\hline
\end{tabular}
\caption{The table summarizes the average precision, recall and F1 scores for planted cliques for both true clique in $\bA^{(c)}$ and pseudo-clique in $\bX^{(c)}$ settings over $nMC = 50$ simulations across different planted clique sizes $n_c$ and different algorithms. }
\label{table:algos}
\end{table}

\newpage
\section{extra figures}

\begin{figure}[!hbp]
\begin{center}
    \subfloat[$n_c = \log(n)$]{\includegraphics[width=.33\textwidth]{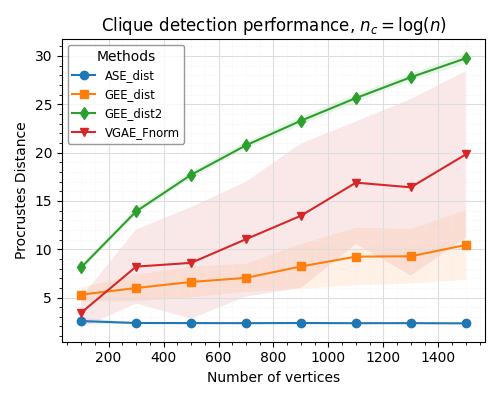}}
    \subfloat[$n_c = \log(n)$]{\includegraphics[width=.33\textwidth]{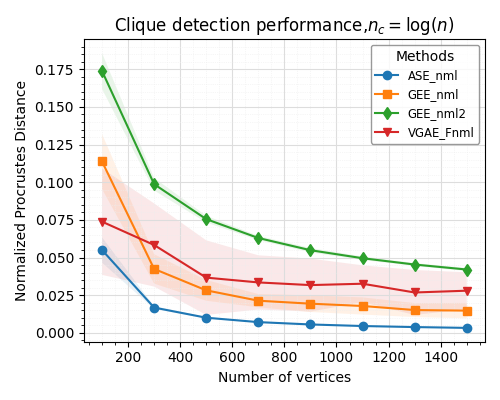}} 
    {\includegraphics[width=.33\textwidth]{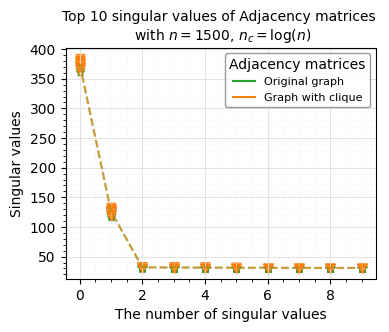}} \\
    \subfloat[$n_c = \log(n), K=3$]{\includegraphics[width=.33\textwidth]{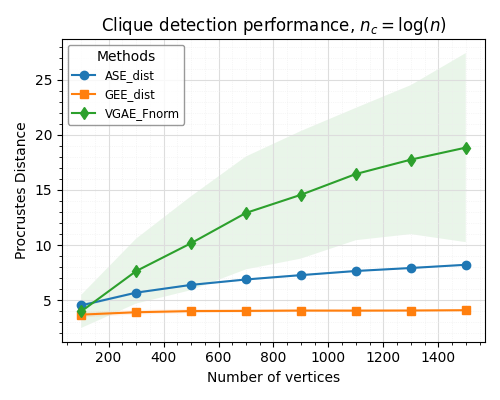}} 
    \subfloat[$n_c = \log(n), K=3$]{\includegraphics[width=.33\textwidth]{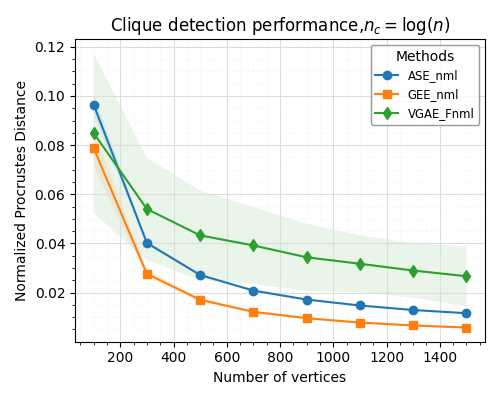}} 
    {\includegraphics[width=.33\textwidth]{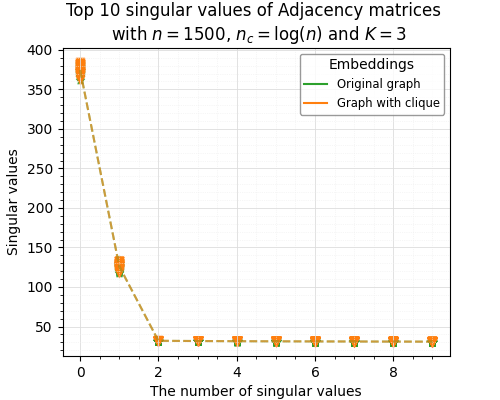}} 
\caption{
For each set of vertices ranges from $n =[100, 300, \dots, 1500] $, we generate a pair of graphs $(G,G^{(c)})$. $G$ is directly sampled from a specified latent position matrix, while $G^{(c)}$ is derived from $G$ by introducing a clique with a size of $n_c = \log(n)$ for the planted pseudo-clique (option ii).  The average distances ($\pm$ 2 s.d.) are displayed in this Figure resulting from $nMC=50$ repetitions.}
\label{fig:logextra1}
\end{center}
\end{figure}

\begin{figure}[!hbp]
\begin{center} 
    \subfloat[$n_c = \log^2(n)$]{\includegraphics[width=.33\textwidth]{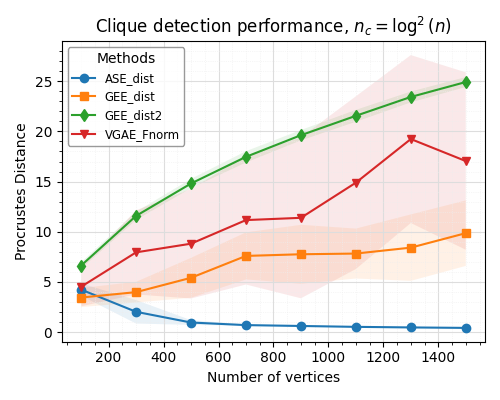}}
    \subfloat[$n_c = \log^2(n)$]{\includegraphics[width=.33\textwidth]{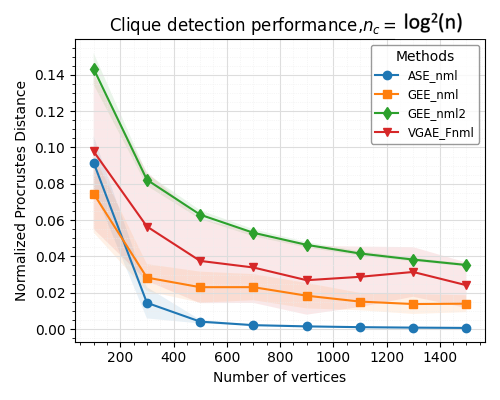}}\\
    \subfloat[$n_c = \log^2(n), K=3$]{\includegraphics[width=.33\textwidth]{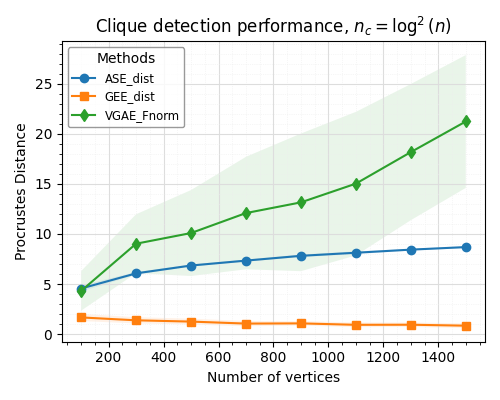}} 
    \subfloat[$n_c = \log^2(n), K=3$]{\includegraphics[width=.33\textwidth]{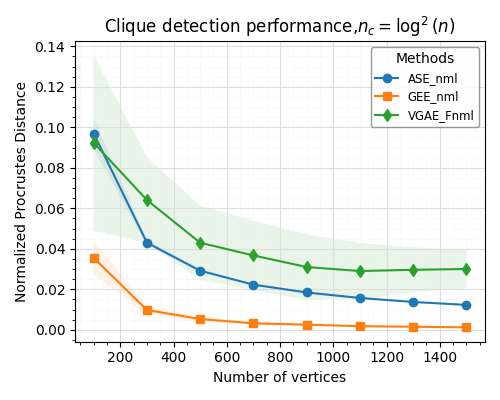}}  
\caption{
For each set of vertices ranges from $n =[100, 300, \dots, 1500] $, we generate a pair of graphs $(G,G^{(c)})$. $G$ is directly sampled from a specified latent position matrix, while $G^{(c)}$ is derived from $G$ by introducing a clique with a size of $n_c = \log^2(n)$ for the planted true-clique (option i).  The average distances ($\pm$ 2 s.d.) are displayed in this Figure resulting from $nMC=50$ repetitions.}
\label{fig:logextra2}
\end{center}
\end{figure}

\end{document}